\documentclass[UTF8,twocolumn,nopacs,floatfix,amsmath,nofootinbib,superscriptaddress,amssymb,preprintnumbers,floatfix]{revtex4}
\usepackage{amsmath}
\usepackage{longtable,lscape,booktabs}
\usepackage{txfonts}
\usepackage{overpic}
\usepackage{amssymb}
\usepackage{color}
\usepackage{longtable,lscape,bm}
\usepackage{multirow}
\usepackage{float}
\usepackage{graphicx,color,dcolumn,bm}
\usepackage{subfigure}
\usepackage{hyperref}
\hypersetup{colorlinks,citecolor=blue,anchorcolor=red,menucolor=red, linkcolor=red,filecolor=red,runcolor=red,urlcolor=blue,frenchlinks=true}
\usepackage{appendix}

\allowdisplaybreaks

\begin{document}

	\title{Anticharmed strange pentaquarks from the one-boson-exchange model}
	
	\author{Nijiati Yalikun}
	\email{nijiati@itp.ac.cn}
	\affiliation{CAS Key Laboratory of Theoretical Physics, Institute
		of Theoretical Physics,  Chinese Academy of Sciences, Beijing 100190, China}
	\affiliation{School of Physical Sciences, University of Chinese Academy of Sciences, Beijing 100049, China}
	
	\author{Bing-Song Zou}\email{zoubs@mail.itp.ac.cn}
	\affiliation{CAS Key Laboratory of Theoretical Physics, Institute
		of Theoretical Physics,  Chinese Academy of Sciences, Beijing 100190, China}
	\affiliation{School of Physical Sciences, University of Chinese Academy of Sciences, Beijing 100049, China}
	\affiliation{School of Physics and Electronics, Central South University, Changsha 410083, China}

\begin{abstract}
Inspired by the LHCb's discovery of hidden-charm pentaquarks, the anticharmed strange pentaquarks $P_{\bar c s}$ in a favor of the hadronic molecule picture are investigated from the one-boson-exchange model. Similar to the hidden-charm pentaquarks, three molecular bound states, one with spin parity $J^P=1/2^-$ below the $\bar D\Sigma$ threshold and another two with $J^P=\{1/2^-,3/2^-\}$ below the $\bar D^*\Sigma$ threshold, are predicted at $3-3.2$ GeV mass region. The mass ordering of the later two states can be interchanged by different reductions of the $\delta(\bm r)$ term. Furthermore, resonances associated with these three bound states are examined by considering $ D_s^-N-\bar D\Lambda-\bar D \Sigma-\bar D^*\Lambda-\bar D^*\Sigma$ coupled-channel dynamics, and decay widths of them are predicted. Our study indicates that the $ D_s^- p$ invariant mass spectrum in the $\bar B_s^0\to \bar n  D_s^- p$ decay is an appropriate place to detect the $P_{\bar c s}$ pentaquarks.  

\end{abstract}

\maketitle

\section{introduction}\label{sec:intro}
 In the past two decades, 
 many candidates of exotic tetraquark and pentaquark states in the charm sector have been observed. An intriguing fact is that most of them are located near some hadron-hadron thresholds. This property can be understood as an attraction between the relevant hadron pair~\cite{Dong:2020hxe}, and naturally leads to the hadronic molecular interpretation for them (see Refs.~\cite{Chen:2016qju,Guo:2017jvc,Brambilla:2019esw,Yamaguchi:2019vea,Dong:2021juy}). The validity of the hadronic molecular picture is also reflected by the successful quantitative predictions of some exotic states in the theoretical works focused on the hadron-hadron interaction~\cite{Tornqvist:1993ng,Wu:2010jy,Wu:2010vk,Wang:2011rga,Yang:2011wz,Wu:2012md,Xiao:2013yca,Uchino:2015uha,Karliner:2015ina}. The LHCb pentaquarks discovered at the recent experiment at the LHC~\cite{Aaij:2015tga,Aaij:2016phn,LHCb:2019kea,LHCb:2021chn} are similar with the hidden-charm $N^*$ states predicted at the mass region above 4.2~GeV~\cite{Wu:2010jy,Wu:2010vk,Wang:2011rga,Yang:2011wz,Wu:2012md,Yuan:2012wz,Xiao:2013yca,Uchino:2015uha,Karliner:2015ina}, and lead to extensive investigation on dynamics of strong interaction between a hadron pair. With the fact that the masses of $P_c(4312)$, $P_c(4440)$ and $P_c(4457)$ are just below the thresholds of the $\Sigma_c \bar{D}$ and $\Sigma_c \bar{D}^*$ hadron pairs, the survey on the hadronic molecules formed by light meson exchange dynamics have been extending our knowledge of exotic states~\cite{Guo:2019kdc,Xiao:2020frg,Guo:2019fdo,He:2019ify,Shimizu:2019jfy,Weng:2019ynv,Voloshin:2019aut,PavonValderrama:2019nbk,Xu:2020gjl,Yamaguchi:2019seo,Ke:2019bkf,Giachino:2020rkj,Yang:2020twg,Azizi:2020ogm,Liu:2020hcv,Phumphan:2021tta,Du:2021fmf,Yalikun:2021bfm}. 

\begin{figure}[ht!]\centering
	\includegraphics[width=0.45\textwidth]{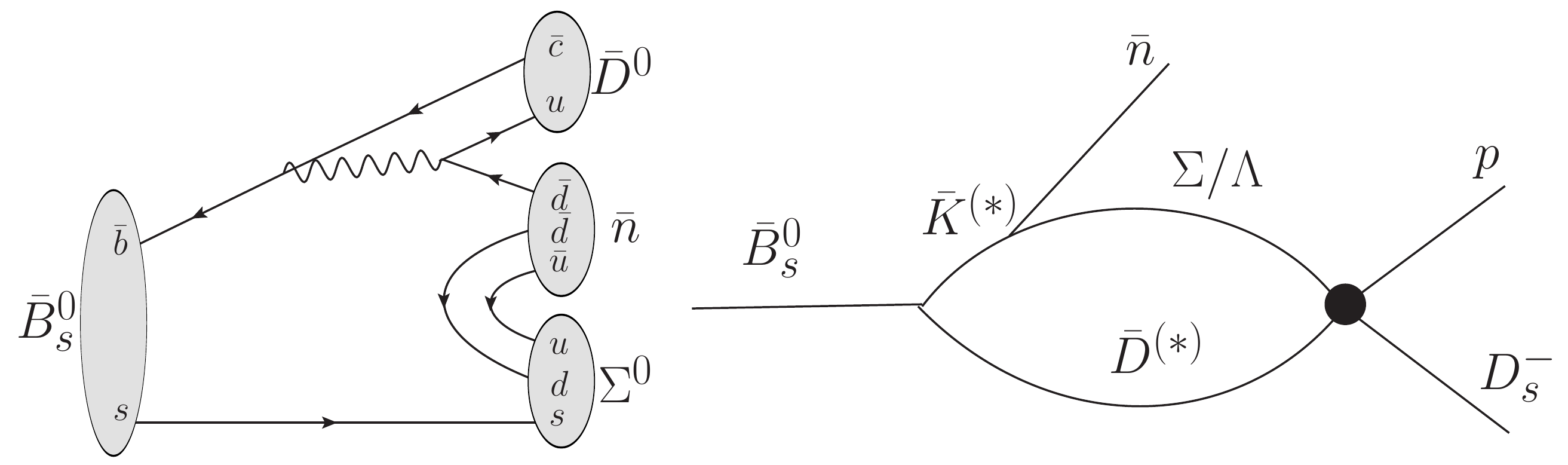}
	\caption{Diagrams showing the production of $P_{\bar c s}$ in $D^-_s p$ final state from $\bar B_s^0$ decay.\label{fig:Pcbs-prod}}
\end{figure}

In the present work, we study in detail the $\bar D^{(*)}\Sigma$ molecular systems within the one boson exchange (OBE) model,  and predict possible anticharmed strange pentaquarks, which can be observed in $ D_s^- p$ invariant mass spectrum in $\bar B_s^0\to  \bar nD_s^- p$ decay. Originally,  a possible anticharmed strange pentaquark $P_{\bar c s}$ as $ D_s^-N$ bound state was proposed in Refs.~\cite{Lipkin:1987sk,Gignoux:1987cn}. A similar bound state was also reported in the coupled-channel $ D_s^- N-\bar D\Lambda-\bar D\Sigma$ system with spin parity $J^P=1/2^-$ and isospin $I=1/2$ in Ref.~\cite{Hofmann:2005sw}. Different from the $ D_s^- N$ bound state, our proposed $\bar D^{(*)}\Sigma$ bound states are well above the $ D_s^- N$ threshold and hence can decay to $ D_s^- p$ to be detected through the $\bar B_s^0\to  \bar nD_s^-p$ process. The corresponding production mechanism of $P_{\bar cs}$ pentaquarks states is illustrated with Fig.~\ref{fig:Pcbs-prod}.   
Recently, an amplitude analysis of $B_s^0\to J/\psi p\bar p$ decay preformed by the LHCb Collaboration indicates the existence of another pentaquark $P_c(4337)$~\cite{LHCb:2021chn},
in which decays of $B_s^0$ or $\bar B_s^0$ are not distinguished and analyzed together. The observation of $P_c(4337)$ stimulates our present study of the $P_{\bar c s}$ pentaquarks which are likely to be observed from the similar $\bar B_s^0$ decay. 

From the aspect of molecular picture, the formation mechanisms of the $P_{\bar cs}$ are similar with the LHCb hidden-charm pentaquarks $P_c(4312)$, $P_c(4440)$ and $P_c(4457)$. A series of work reveals the molecular nature of these pentaquarks. For instance, $P_c(4312)$, $P_c(4440)$ and $P_c(4457)$ are well reproduced as three isodoublet states near the $\bar D\Sigma_c$ and $\bar D^*\Sigma_c$ threshold with spin parity $J^P=1/2^-$ and $J^P=\{1/2^-,3/2^-\}$~\cite{He:2019rva,Chen:2019asm,Liu:2019zvb,Du:2021fmf}. The spin parities of the $P_c(4440)$ and $P_c(4457)$ pentaquarks can be interchanged by the mechanism induced by the parametrized $\delta(\bm r)$ term (it is related to short-range physics)~\cite{Yalikun:2021bfm}. Somehow, there is an analogy between $\bar D^{(*)}\Sigma_c$ and $\bar D^{(*)}\Sigma$ systems, in which the charm quark in the $\Sigma_c$ component of the former hadron pair is replaced by a strange quark. Hence, it comes to our mind to study whether the similar pattern as the LHCb hidden-charm pentaquarks exists or not. Similar molecular systems $D^{(*)}\Sigma$, are studied in Ref.~\cite{Yu:2018yxl} with a chiral unitary approach, and they are used to explain the molecular nature of $\Xi_c(2970)$ and $\Xi_c(3055)$ states. To understand the structure of $\Lambda_c(2940)$, the molecular systems $D^{(*)}N$ are also studied with the OBE model in Refs.~\cite{He:2006is,He:2010zq}, and with the chiral effective field theory to the next-to-leading order in Ref.~\cite{Wang:2020dhf}. The $D\Xi$ system, SU(3) partner of the $D^{(*)}N$ and $D^{(*)}\Sigma$, is studied with the OBE model, and several molecular pentaquarks in this sector are explained as an $\Omega_c^*$ state~\cite{Huang:2018wgr}. However, the detailed investigation on the hadron pairs in the anticharm sector, such as $\bar D^{(*)}N$, $\bar D^{(*)}\Sigma$ and $\bar D^{(*)}\Xi$, is  still inadequate. 
 
In this work, the molecular systems $\bar D^{(*)}\Sigma$ are investigated in the OBE model using the effective Lagrangian approach. The coupled-channel dynamics of $ D_s^-N-\bar D\Lambda-\bar D \Sigma-\bar D^*\Lambda-\bar D^*\Sigma$ are considered to evaluate the decay width of $P_{\bar cs}$, $S-D$ wave mixing effects are also included. The OBE potentials for various channels are derived from the effective Lagrangian, taking into account of heavy quark spin symmetry (HQSS)~\cite{Yan:1992gz,Wise:1992hn,Cho:1994vg,Casalbuoni:1996pg} and  SU(3)  flavor symmetry~\cite{Erkelenz:1974uj,Doring:2010ap,deSwart:1963pdg}. The possible bound states in the single channel interaction, and resonances associated with these bound states are obtained by means of solving the Schr\"odinger equation. Also, partial widths of these resonances decaying to lower channels are evaluated.

\section{potentials}\label{sec:potentials}
The OBE potential model for the pair of hadrons is quite successful in interpreting the formation mechanisms of pentaquarks~\cite{He:2019rva,Chen:2019asm,Liu:2019zvb,Du:2021fmf,Yalikun:2021bfm}.
 In this work, we also use the OBE potentials of  $D_s^-N$, $\bar D\Lambda$, $\bar D \Sigma$, $\bar D^*\Lambda$ and $\bar D^*\Sigma$ systems to investigate the possible $P_{\bar c s}$ pentaquarks.
The interactions of anticharmed meson with light scalar, pseudoscalar and vector mesons can be described by the effective Lagrangian, taking into account of HQSS and SU(3) flavor symmetry~\cite{Cheng:1992xi,Yan:1992gz,Wise:1992hn,Cho:1994vg,Casalbuoni:1996pg,Pirjol:1997nh,Liu:2011xc}. The effective vertices relevant to our work are   
\begin{align}
\mathcal{L}_{\tilde{\mathcal{P}}\tilde{\mathcal{P}}\sigma}&=2g_S\tilde{\mathcal{P}}^{*\mu\dagger}_{a}\sigma \tilde{\mathcal{P}}^*_{a\mu}-2g_S\tilde{\mathcal{P}}^\dagger_a \sigma \tilde{\mathcal{P}}_a,\label{lag:ppsigma}\\
\mathcal{L}_{\tilde{\mathcal{P}}\tilde{\mathcal{P}} V}&=-\sqrt{2}\beta g_V\tilde{\mathcal{P}}^{*\dagger}_{a\mu}v_\alpha \mathbb{V}^\alpha_{ab}\tilde{\mathcal{P}}^{*\mu}_b-i2\sqrt{2}\lambda g_V \tilde{\mathcal{P}}^{*\dagger}_{a\mu}F^{\mu\nu}_{ab}(\mathbb{V})\tilde{\mathcal{P}}^*_{b\nu}\nonumber\\
	& +(2\sqrt{2}\lambda g_V \varepsilon^{\alpha\beta\mu\kappa}v_\kappa\tilde{\mathcal{P}}^{*\dagger}_{a\mu}\partial_\alpha\mathbb{V}_{ab\beta}\tilde{\mathcal{P}}_b+H.c.)\notag\\
	& +\sqrt{2}\beta g_V\tilde{\mathcal{P}}^\dagger_a v_\alpha\mathbb{V}^\alpha_{ab} \tilde{\mathcal{P}}_b,\label{lag:ppV}\\
\mathcal{L}_{\tilde{\mathcal{P}}\tilde{\mathcal{P}}\mathbb{P}}&=i\frac{2g}{f_\pi}\varepsilon^{\alpha\mu\nu\kappa} v_\kappa \tilde{\mathcal{P}}^{*\dagger}_{a\mu}\partial_\alpha\mathbb{P}_{ab}\tilde{\mathcal{P}}^*_{b\nu}+\frac{2g}{f_\pi}(\tilde{\mathcal{P}}^{*\dagger}_{a\mu}\partial^\mu\mathbb{P}_{ab}\tilde{\mathcal{P}}_b+H.c.),\label{lag:ppP}
\end{align}
where flavor indices are denoted by $a$ and $b$. The anticharmed meson fields represented by scaled fields $\tilde{\mathcal{P}}^{(*)}$ are defined in flavor/isospin space as $(\bar D^0,D^-, D_s^{-} )$ and $(\bar D^{*0},D^{*-}, D_s^{*-})$. The $\sigma$ is the lightest scalar meson and its physics is governed by the dynamics of the Goldstone bosons and relevant to the interaction between two pions~\cite{Bardeen:2003kt,Machleidt:1987hj}.
   The light pseudoscalar octet and the vector nonet are denoted by $\mathbb{P}$ and $\mathbb{V}^\alpha$, respectively~\cite{Yalikun:2021bfm}. The $\eta$ meson in the $\mathbb{P}$ is identified as $\eta_8$ state, while the $\omega$ and $\phi$ in $\mathbb{V}^\alpha$ are represented by ideal mixing of $\omega_8$ and $\omega_1$ states. Field strength tensor is $F^{\mu\nu}_{ab}(\mathbb{V})=(\partial^\mu \mathbb{V}^\nu-\partial^\nu \mathbb{V}^\mu+i\frac{g_V}{\sqrt 2}[\mathbb{V}^\mu,\mathbb{V}^\nu]_-)_{ab}$, where $[A,B]_-=AB-BA$. The scalar meson coupling $g_S$ is taken as $g_S=g_\pi/(2\sqrt{6})$ with $g_\pi=3.73$~\cite{Bardeen:2003kt}. The pseudoscalar meson coupling is taken as $g=-0.59$, which is extracted from the experimental decay widths of $D^*\to D \pi $~\cite{ParticleDataGroup:2020ssz} with the pion decay constant $f_\pi=132$ MeV, and the sign is determined from the quark model~\cite{Meng:2019ilv}. Vector meson couplings are taken as $g_V=5.9$, $\beta=0.9$ and $\lambda=0.56~\rm{GeV}^{-1}$, in which $g_V$ and $\beta$ are determined by vector meson dominance~\cite{Bando:1987br}, $\lambda$ is obtained by comparing the form factor calculated by light cone sum rule with that obtained by lattice calculation~\cite{Isola:2003fh}. These couplings are also used in recent studies focusing on molecular $P_c$ pentaquarks~\cite{He:2019rva,He:2019ify,Chen:2019asm,Yalikun:2021bfm}. On the other hand, the effective Lagrangian depicting the interaction of the baryons in the SU(3) flavor octet ($\mathcal{B}$) with light scalar, pseudoscalar and vector mesons reads
\footnote{ In these Lagrangians, flavor information of the particles in the SU(3) multiplet is embedded in the coupling constants $g_{\mathcal{B} \mathcal{B}\sigma}$, $g_{\mathcal{B} \mathcal{B}P}$, $g_{\mathcal{B} \mathcal{B}V}$ and $\kappa_{\mathcal{B} \mathcal{B}V}$. Explicit form of these Lagrangians can be found in Appendix~\ref{appen:lag}. In actual calculation, we consider small SU(3) flavor symmetry breaking effects by setting $m_{\mathcal{B}}$ and $m_P$ to the physical masses of particles in octet baryon and pseudoscalar meson matrices, respectively.}
~\cite{Erkelenz:1974uj,Doring:2010ap,Schutz:1998jx,Krehl:1999km,Bryan:1969mp,Ronchen:2012eg,Liu:2011xc,Holzenkamp:1989tq}. 
	\begin{align}
\mathcal{L}_{\mathcal{B} \mathcal{B} \sigma}&=-g_{\mathcal{B} \mathcal{B}\sigma}\bar\psi\phi_\sigma \psi,\label{lag:BBsigma}\\
\mathcal{L}_{\mathcal{B} \mathcal{B} V}&=-g_{\mathcal{B} \mathcal{B} V}\bar\psi\{ \gamma_\mu-\frac{\kappa_{\mathcal{B} \mathcal{B} V}}{2 m_{\mathcal{B}}}\sigma_{\mu\nu}\partial^\nu\}V^\mu\psi,\label{lag:BBV}\\
\mathcal{L}_{\mathcal{B} \mathcal{B} P}&=-\frac{g_{\mathcal{B} \mathcal{B}P}}{m_P}\bar\psi \gamma^5\gamma_\mu \partial^\mu\phi_{P}\psi,\label{lag:BBP}
\end{align}
where $\psi$ represents the Dirac field operator for the SU(3) octet baryon, $\phi_\sigma$, $V^\mu$ and $\phi_{P}$ are field operators corresponding to the  scalar, vector and pseudoscalar mesons, respectively.
 The scalar meson couplings for the octet baryons are given in Refs.~\cite{Holzenkamp:1989tq,Ronchen:2012eg} as $g_{NN\sigma}=8.465$, $g_{\Lambda\Lambda\sigma}=7.579$ and $g_{\Sigma\Sigma\sigma}=10.85$. The pseudoscalar and vector couplings for nucleon $N$ are also available in Ref.~\cite{Ronchen:2012eg} as $g_{NN\rho}=3.25$, $g_{NN\pi}=0.989$, $\kappa_{NN\rho}=6.1$ and $\kappa_{NN\omega}=0$. These couplings are determined by a fit to the empirical hyperon-nucleon ($\Lambda N,\Sigma N$) data. {\color{blue} The relative signs of them with respect to the Lagrangian in Eqs.~\eqref{lag:ppsigma}-\eqref{lag:ppP} are fixed by the quark model (see Appendix~\ref{appen:phase}) and are consistent with earlier corresponding studies for the $P_c$ states~\cite{Wu:2010jy,Chen:2019asm,Liu:2019zvb,Yalikun:2021bfm}}. The pseudoscalar and vector meson couplings for the other octet baryons ($\Sigma,\Lambda$ and $\Xi$) are obtained by means of SU(3) flavor symmetry~\cite{deSwart:1963pdg}.

\begin{figure*}[ht!]\centering
	\subfigure[]{\label{fig:feyn-type1}\includegraphics[width=0.25\textwidth]{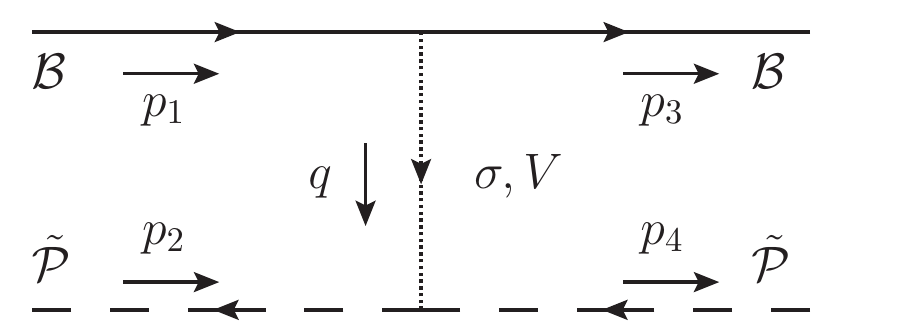}}	
	\subfigure[]{\label{fig:feyn-type1}\includegraphics[width=0.25\textwidth]{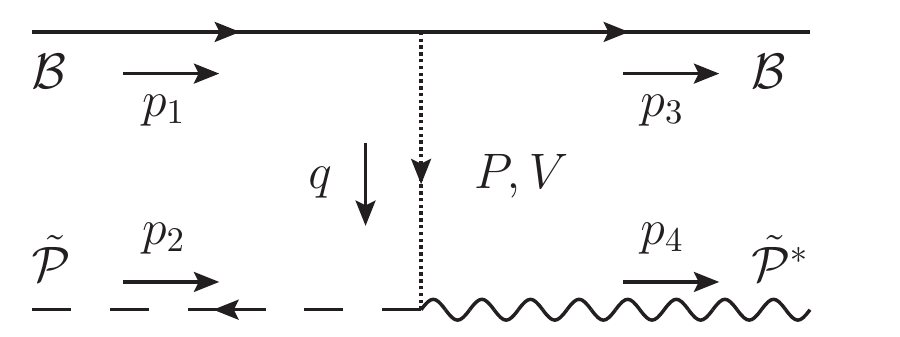}}	
	\subfigure[]{\label{fig:feyn-type1}\includegraphics[width=0.25\textwidth]{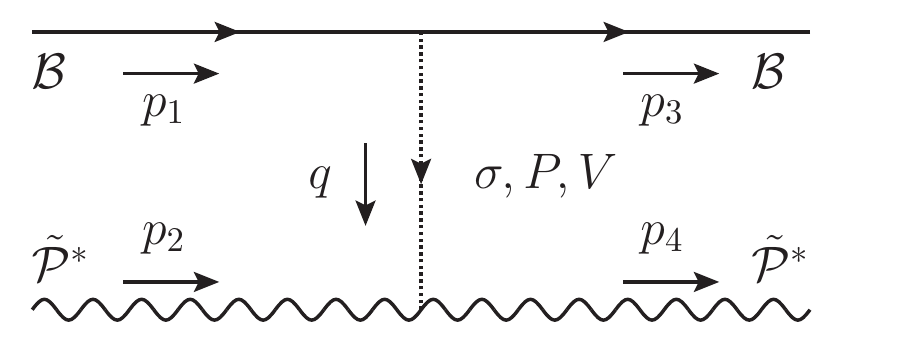}}
	\caption{The $t$-channel Feynman diagram for the processes (a) $\tilde{\mathcal{P}}\mathcal{B}\to \tilde{\mathcal{P}}\mathcal{B}$, (b) $\tilde{\mathcal{P}}\mathcal{B}\to \tilde{\mathcal{P}}^{*}\mathcal{B}$ and (c) $\tilde{\mathcal{P}}^{*}\mathcal{B}\to \tilde{\mathcal{P}}^{*}\mathcal{B}$. The exchanged scalar, vector and pseudoscalar mesons are labeled with $\sigma$, $V$ and $P$, respectively.    }\label{fig:Feyn}
\end{figure*}

The potentials for the $\tilde{\mathcal{P}}^{(*)}\mathcal{B}$ system in the momentum space can be obtained from the $t$-channel scattering amplitudes ($\mathcal{M}$) with the Breit approximation~\cite{Wang:2020dya}
\begin{align}
	\mathcal{V}^{h_1h_2\to h_3h_4}(\bm q)=-\frac{\mathcal{M}^{h_1h_2\to h_3h_4}}{\sqrt{2m_1m_22m_3m_4}},\label{eq:Breit}
\end{align}
 where $m_1$ and $m_2$ are the masses of the particles $h_1$ and $h_2$ in the initial state, while $m_3$ and $m_4$ are the masses of particles $h_3$ and $h_4$ in the final state. The $t$-channel Feynman diagrams for the scattering processes $\tilde{\mathcal{P}}\mathcal{B}\to \tilde{\mathcal{P}}\mathcal{B}$, $\tilde{\mathcal{P}}\mathcal{B}\to \tilde{\mathcal{P}}^{*}\mathcal{B}$ and $\tilde{\mathcal{P}}^{*}\mathcal{B}\to \tilde{\mathcal{P}}^{*}\mathcal{B}$ are shown in Fig.~\ref{fig:Feyn}.  
 Their amplitudes are calculated with the Lagrangian given in Eqs.~\eqref{lag:ppsigma}$-$\eqref{lag:BBP}. In our calculation, we use the Dirac spinor with positive energy for $\mathcal{B}$ as  
\begin{align}
	u(\bm p,s)=\sqrt{E+M}\begin{pmatrix}
		\chi\\
		\frac{\sigma \cdot \bm p}{E+M}\chi
	\end{pmatrix},
\end{align}
which is normalized to $2M$, where $\sigma$ is Pauli matrix, and $\chi$ is two components spinor. For the scaled-heavy meson fields $\tilde{\mathcal{P}}$ and $\tilde{\mathcal{P}}^*$, we adopt the normalization relations as
$\langle 0 |\tilde{\mathcal{P}}| \bar c q(0^-)\rangle=\sqrt{M_{\tilde{\mathcal{P}}}} $ and $\langle 0|\tilde{\mathcal{P}}^*_\mu|\bar c q(1^-)\rangle=\epsilon_\mu\sqrt{M_{\tilde{\mathcal{P}}^*}} $~\cite{Wise:1992hn,Wang:2020dya}. In the center-of-mass frame, the four-momenta of the particles in the initial state are $p_1=(E_1,\bm p)$ and $p_2=(E_2,-\bm p)$, while the four-momenta of the particles in the final state are $p_3=(E_3,\bm p')$ and $p_4=(E_4,-\bm p')$. The four-momentum of the exchanged meson is given by $q=p_3-p_1=p_2-p_4=(q^0,\bm q)$. For the convenience of the calculation, we define the new variables  
\begin{align}
	\bm q=\bm p'-\bm p,\ \bm Q=\frac{1}{2}(\bm p' +\bm p).
\end{align}     
With Breit approximation in Eq.~\eqref{eq:Breit}, the potentials in the momentum space for the scattering processes $\tilde{\mathcal{P}}\mathcal{B}\to \tilde{\mathcal{P}}\mathcal{B}$, $\tilde{\mathcal{P}}\mathcal{B}\to \tilde{\mathcal{P}}^{*}\mathcal{B}$ and $\tilde{\mathcal{P}}^{*}\mathcal{B}\to \tilde{\mathcal{P}}^{*}\mathcal{B}$ are derived keeping up to $1/m_{\mathcal{B}}^2$ order and listed as the three types in Eq.~\eqref{eq:poten-in-p}.
\begin{widetext}
\begin{subequations}\label{eq:poten-in-p}
	\begin{itemize}
		\item Type I: $\tilde{\mathcal{P}}\mathcal{B}\to \tilde{\mathcal{P}}\mathcal{B}$
		\begin{align}
			\mathcal{V}_\sigma^I(\bm q,\bm Q) &= -\tau_\sigma g_{\mathcal{B}\mathcal{B}\sigma}g_S [ 1-\frac{i\bm\sigma\cdot(\bm q\times\bm Q)}{4 m_{\mathcal{B}}^2}] \frac{1}{\bm q^2+\mu_\sigma^2},\\
			\mathcal{V}_V^I(\bm q,\bm Q) &=-\tau_V\frac{g_{\mathcal{B}\mathcal{B} V}\beta g_V}{2} [ 1+\frac{1+2\kappa_{\mathcal{B}\mathcal{B}V}}{4m_{\mathcal{B}}^2}i\bm\sigma\cdot(\bm q \times \bm Q)] \frac{1}{\bm q^2+\mu_V^2},
			\end{align}
		\item Type II: $\tilde{\mathcal{P}}\mathcal{B}\to \tilde{\mathcal{P}^*}\mathcal{B}$
		\begin{align}
			\mathcal{V}_{P}^{II}(\bm q,\bm Q) &= \tau_P\frac{gg_{\mathcal{B}\mathcal{B}\pi}}{\sqrt 2 f_\pi m_P} \frac{\bm\sigma\cdot \bm q \bm \epsilon_4^*\cdot \bm q}{\bm q^2+\mu_{P}^2} ,\\
			\mathcal{V}_V^{II}(\bm q,\bm Q) &=\tau_V\frac{g_{\mathcal{B}\mathcal{B} V}\lambda g_V}{2m_{\mathcal{B}}}  [(2+3\kappa_{\mathcal{B}\mathcal{B}V})i\bm\epsilon_4^*\cdot (\bm q\times \bm Q) +(1+\kappa_{\mathcal{B}\mathcal{B}V})(\bm q^2\bm\epsilon_4^*\cdot \bm\sigma -\bm\sigma\cdot \bm q \bm\epsilon_4^*\cdot \bm q) ] \frac{1}{\bm q^2+\mu_V^2},
			\end{align}
		\item Type III: $\tilde{\mathcal{P}}^*\mathcal{B}\to \tilde{\mathcal{P}}^*\mathcal{B}$
		\begin{align}
			\mathcal{V}_\sigma^{III}(\bm q,\bm Q) &= -\tau_\sigma g_{\mathcal{B}\mathcal{B}\sigma}g_S\bm{\epsilon}^*_4 \cdot \bm{\epsilon}_2 [ 1-\frac{i\bm\sigma\cdot(\bm q\times\bm Q)}{4 m_{\mathcal{B}}^2} ] \frac{1}{\bm q^2+\mu_\sigma^2},\\
			\mathcal{V}_{P}^{III}(\bm q,\bm Q) &= \tau_P\frac{gg_{\mathcal{B}\mathcal{B}\pi}}{\sqrt 2 f_\pi m_P} \frac{\bm\sigma\cdot \bm q \mathcal{T}\cdot \bm q}{\bm q^2+\mu_{P}^2} ,\\
			\mathcal{V}_V^{III}(\bm q,\bm Q) &=-\tau_V \frac{g_{\mathcal{B}\mathcal{B} V}\beta g_V}{ 2} [ 1+\frac{\kappa_{\mathcal{B}\mathcal{B}V}}{2m_{\mathcal{B}}^2}i\bm\sigma\cdot(\bm q \times \bm Q)] \frac{\bm{\epsilon}^*_4 \cdot \bm{\epsilon}_2}{\bm q^2+\mu_V^2}\nonumber\\
			&\, -\tau_V\frac{g_{\mathcal{B}\mathcal{B} V}\lambda g_V }{ 2m_{\mathcal{B}}}  [(2-\kappa_{\mathcal{B}\mathcal{B}V})i\mathcal{T}\cdot (\bm q\times \bm Q) -(1+\kappa_{\mathcal{B}\mathcal{B}V})(\bm\sigma\times\bm q)\cdot(\mathcal{T}\times \bm q) ] \frac{1}{\bm q^2+\mu_V^2}\label{eq:VIII-in-p},
			\end{align}
	\end{itemize}
	\end{subequations}
\end{widetext}
where the subindices $V$ and $P$ represent the exchanged vector and  pseudoscalar mesons, respectively; The polarization vectors for the $\tilde{\mathcal{P}}^*$ meson at the final and initial states are denoted by $\bm\epsilon_4^*$ and $\bm\epsilon_2$, respectively; $\mathcal{T}$ is defined as $\mathcal{T}=i\bm\epsilon_2\times\bm\epsilon_4^*$. In the inelastic scattering, the energy of the exchanged meson is nonzero, so the denominator of the propagator can be rewritten as $q^2-m^2_{\rm{ex}}=(q^0)^2-\bm q^2-m_{\rm{ex}}^2=-(\bm q^2+\mu^2_{\rm{ex}})$, where $\mu_{\rm{ex}}$ represents the effective mass of the exchanged meson. In the center-of-mass frame, the energy of the exchanged meson, $q^0$, is calculated as 
\begin{align}
	q^0=\frac{m_2^2-m_1^2+m_3^2-m_4^2}{2(m_3+m_4)},
\end{align} 
where $m_1$ and $m_2$ are the masses of the particles in the initial state, while $m_3$ and $m_4$ are the masses of particles in the final state. 
The coupling constants and the isospin factors for the potentials of the specific scattering processes are listed in Table~\ref{tab:coupling}, where the relations of the coupling constants in the  SU(3)  flavor symmetry are adopted~\cite{deSwart:1963pdg,Ronchen:2012eg}. The isospin factors of each meson exchange potential for the total isospin $I=1/2$ system are listed in the column labeled with $\tau_{\rm{ex}}$. The potentials for the specific scattering process listed in the column labeled with ``Transition" in Table~\ref{tab:coupling} are obtained from the corresponding type of the potentials in Eq.~\eqref{eq:poten-in-p} by replacing the coupling constants. For instance, the $\rho$ meson exchange potential for the process $\bar D^*\Sigma \to \bar D^*\Sigma$ is obtained by replacing the coupling constants $g_{\mathcal{B}\mathcal{B}V}$ and $\kappa_{\mathcal{B}\mathcal{B}V}$ in Eq.~\eqref{eq:VIII-in-p} with $2g_{NN\rho}\alpha_V$ and $\frac{\kappa_{NN\rho}}{4\alpha_V}$, respectively.     
	 \begin{table}[ht!]\centering
		\caption{The type of potential, exchanged meson, isospin factor for $I=1/2$ and coupling constant for specific channel. The label ``$\cdots$'' means that the coupling is forbidden. The values of the mixing parameters $\alpha_P$ and $\alpha_V$ are taken from Ref.~\cite{Ronchen:2012eg}, namely $\alpha_P=0.4$ and $\alpha_V=1.15$, which connect the couplings of the nucleon to other particles in the octet baryon matrix.}\label{tab:coupling}
		\begin{ruledtabular}
		\begin{tabular}{cccccc}
	Transition&Type&Ex.&$\tau_{\rm{ex}}$&$g_{\mathcal{B}\mathcal{B}\rm{ex}}$&$\kappa_{\mathcal{B}\mathcal{B}\rm{ex}}$\\
	\hline
	$\bar D^*\Sigma\to\bar D^*\Sigma$&III&$ \begin{matrix} \sigma \\ \rho \\ \omega \\ \pi\\ \eta \end{matrix} $&$ \begin{matrix} 1 \\ -2 \\ 1 \\ -2\\ \frac{1}{\sqrt 3} \end{matrix} $&$ \begin{matrix} g_{\Sigma\Sigma\sigma} \\ 2g_{NN\rho}\alpha_V \\ 2g_{NN\rho}\alpha_V \\ 2g_{NN\pi}\alpha_P\\ \frac{2}{\sqrt 3}g_{NN\pi}(1-\alpha_P) \end{matrix} $&$ \begin{matrix}\cdots\\ \frac{\kappa_{NN\rho}}{4\alpha_V} \\ \frac{\kappa_{NN\rho}}{4\alpha_V} \\\cdots\\\cdots\end{matrix} $\\
	\hline
	$\bar D^*\Lambda\to\bar D^*\Sigma$&III&$ \begin{matrix} \rho\\ \pi\end{matrix} $&$ \begin{matrix} \sqrt{3}\\\sqrt{3} \end{matrix} $&$ \begin{matrix}\frac{2}{\sqrt{3}}g_{NN\rho}(1-\alpha_V) \\ \frac{2}{\sqrt{3}}g_{NN\pi}(1-\alpha_P)\end{matrix}$&$ \begin{matrix}\frac{3\kappa_{NN\rho}}{4(1-\alpha_V)} \\ \cdots \end{matrix}$\\
	\hline
	$\bar D\Sigma\to\bar D^*\Sigma$&II&$ \begin{matrix}  \rho \\ \omega \\ \pi\\ \eta \end{matrix} $&$ \begin{matrix} -2 \\ 1 \\ -2\\ \frac{1}{\sqrt 3} \end{matrix} $&$ \begin{matrix}  2g_{NN\rho}\alpha_V \\ 2g_{NN\rho}\alpha_V \\ 2g_{NN\pi}\alpha_P\\ \frac{2}{\sqrt 3}g_{NN\pi}(1-\alpha_P) \end{matrix} $&$ \begin{matrix}  \frac{\kappa_{NN\rho}}{4\alpha_V} \\ \frac{\kappa_{NN\rho}}{4\alpha_V} \\\cdots\\\cdots\end{matrix} $\\
	\hline
	$\bar D\Lambda\to\bar D^*\Sigma$&II&$ \begin{matrix}  \rho  \\ \pi \end{matrix} $&$ \begin{matrix}  \sqrt{3}\\\sqrt{3}\end{matrix} $&$ \begin{matrix}\frac{2}{\sqrt{3}}g_{NN\rho}(1-\alpha_V) \\ \frac{2}{\sqrt{3}}g_{NN\pi}(1-\alpha_P)\end{matrix}$&$ \begin{matrix}\frac{3\kappa_{NN\rho}}{4(1-\alpha_V)} \\ \cdots \end{matrix}$\\
	\hline
	$  D_s^-N\to\bar D^*\Sigma$&II&$ \begin{matrix}  \bar K^*  \\ \bar K \end{matrix} $&$ \begin{matrix}  \sqrt{6}\\\sqrt{6}\end{matrix} $&$ \begin{matrix}g_{NN\rho}(1-2\alpha_V) \\ g_{NN\pi}(1-2\alpha_P)\end{matrix}$&$ \begin{matrix}\frac{\kappa_{NN\rho}}{2(1-2\alpha_V)} \\ \cdots \end{matrix}$\\
	\hline
	$\bar D^*\Lambda\to\bar D^*\Lambda$&III&$ \begin{matrix} \sigma\\ \omega \\ \eta\end{matrix} $&$ \begin{matrix} 1 \\ 1 \\ \frac{1}{\sqrt 3} \end{matrix} $&$ \begin{matrix}g_{\Lambda\Lambda\sigma} \\ \frac{2}{3}g_{NN\rho}(5\alpha_V-2)\\ -\frac{2}{\sqrt 3}g_{NN\pi}(1-\alpha_P) \end{matrix}$&$ \begin{matrix} \cdots \\ -\frac{3\kappa_{NN\rho}}{4(5\alpha_V-2)}   \\-  \end{matrix}$\\
	\hline
	$\bar D\Sigma\to\bar D^*\Lambda$&II&$ \begin{matrix}  \rho  \\ \pi \end{matrix} $&$ \begin{matrix}  \sqrt{3}\\\sqrt{3}\end{matrix} $&$ \begin{matrix}\frac{2}{\sqrt{3}}g_{NN\rho}(1-\alpha_V) \\ \frac{2}{\sqrt{3}}g_{NN\pi}(1-\alpha_P)\end{matrix}$&$ \begin{matrix}\frac{3\kappa_{NN\rho}}{4(1-\alpha_V)} \\ \cdots \end{matrix}$\\
	\hline
	$\bar D\Lambda\to\bar D^*\Lambda$&II&$ \begin{matrix} \omega \\ \eta\end{matrix} $&$ \begin{matrix}  1 \\ \frac{1}{\sqrt 3} \end{matrix} $&$ \begin{matrix} \frac{2}{3}g_{NN\rho}(5\alpha_V-2)\\ -\frac{2}{\sqrt 3}g_{NN\pi}(1-\alpha_P) \end{matrix}$&$ \begin{matrix}-\frac{3\kappa_{NN\rho}}{4(5\alpha_V-2)}   \\-  \end{matrix}$\\
	\hline
	$  D_s^-N\to\bar D^*\Lambda$&II&$ \begin{matrix}  \bar K^*  \\ \bar K \end{matrix} $&$ \begin{matrix}  \sqrt{2}\\\sqrt{2}\end{matrix} $&$ \begin{matrix}-\frac{1}{\sqrt 3}g_{NN\rho}(1+2\alpha_V) \\ -\frac{1}{\sqrt 3}g_{NN\pi}(1+2\alpha_P)\end{matrix}$&$ \begin{matrix}\frac{3\kappa_{NN\rho}}{2(1+2\alpha_V)} \\ \cdots \end{matrix}$\\
	\hline
	$\bar D\Sigma\to\bar D\Sigma$&I&$ \begin{matrix} \sigma \\ \rho \\ \omega \end{matrix} $&$ \begin{matrix} 1 \\ -2 \\ 1  \end{matrix} $&$ \begin{matrix} g_{\Sigma\Sigma\sigma} \\ 2g_{NN\rho}\alpha_V \\ 2g_{NN\rho}\alpha_V  \end{matrix} $&$ \begin{matrix}\cdots\\ \frac{\kappa_{NN\rho}}{4\alpha_V} \\ \frac{\kappa_{NN\rho}}{4\alpha_V} \end{matrix} $\\
	\hline
	$\bar D\Lambda\to\bar D\Sigma$&I&$ \begin{matrix} \rho\end{matrix} $&$ \begin{matrix} \sqrt{3} \end{matrix} $&$ \begin{matrix}\frac{2}{\sqrt{3}}g_{NN\rho}(1-\alpha_V) \end{matrix}$&$ \begin{matrix}\frac{3\kappa_{NN\rho}}{4(1-\alpha_V)}  \end{matrix}$\\
	\hline
	$  D_s^-N\to\bar D\Sigma$&I&$ \begin{matrix}  \bar K^*  \end{matrix} $&$ \begin{matrix}  \sqrt{6}\end{matrix} $&$ \begin{matrix}g_{NN\rho}(1-2\alpha_V) \end{matrix}$&$ \begin{matrix}\frac{\kappa_{NN\rho}}{2(1-2\alpha_V)}\end{matrix}$\\
	\hline
	$\bar D\Lambda\to\bar D\Lambda$&I&$ \begin{matrix} \sigma\\ \omega \end{matrix} $&$ \begin{matrix} 1 \\ 1  \end{matrix} $&$ \begin{matrix}g_{\Lambda\Lambda\sigma} \\ \frac{2}{3}g_{NN\rho}(5\alpha_V-2)\end{matrix}$&$ \begin{matrix} \cdots \\ -\frac{3\kappa_{NN\rho}}{4(5\alpha_V-2)} \end{matrix}$\\
	\hline
	$  D_s^-N\to\bar D\Lambda$&I&$ \begin{matrix}  \bar K^*  \end{matrix} $&$ \begin{matrix}  \sqrt{2}\end{matrix} $&$ \begin{matrix}-\frac{1}{\sqrt 3}g_{NN\rho}(1+2\alpha_V) \end{matrix}$&$ \begin{matrix}\frac{3\kappa_{NN\rho}}{2(1+2\alpha_V)}  \end{matrix}$\\
	\hline
	$  D_s^-N\to  D_s^-N$&I&$ \begin{matrix} \sigma\end{matrix} $&$ \begin{matrix} 1  \end{matrix} $&$ \begin{matrix}g_{NN\sigma}\end{matrix}$&$ \begin{matrix} \cdots  \end{matrix}$
		\end{tabular}
	\end{ruledtabular}
	\end{table}

The coordinate space representations of the potentials in Eq.~\eqref{eq:poten-in-p} are obtained by preforming the following Fourier transformation analytically 
\begin{align}
	\mathcal{V}_{\rm{ex}}^{(I,II,III)}=\frac{1}{(2\pi)^3}\int \mathcal{V}_{\rm{ex}}^{(I,II,III)}(\bm q,\bm Q)F^2(\bm q,\tilde{\Lambda},\mu_{\rm{ex}}){\rm{e}}^{i\bm q \cdot \bm r}d^3\bm q,\label{eq:Fourier-trans}
\end{align} 
where the cutoff parameter $\Lambda$ is introduced with the form factor to reflect the inner structure of the interacting vertices~\cite{Tornqvist:1993ng},
\begin{align}
F(\bm q,\tilde{\Lambda},\mu_{\rm{ex}})=\frac{m_{\rm{ex}}^2-\Lambda^2}{(q^0)^2-\bm q^2-\Lambda^2}=\frac{\tilde{\Lambda}^2-\mu_{\rm{ex}}^2}{\bm q^2+\tilde{\Lambda}^2},\label{eq:form-factor}
\end{align}
where we define $\tilde\Lambda=\sqrt{\Lambda^2-(q^0)^2}$ and $\mu_{\rm{ex}}=\sqrt{m_{\rm{ex}}^2-(q^0)^2}$ for convenience. Before giving the potentials in the coordinate space, first, let us fucus on the Fourier transformation of the typical functions in Eq.~\eqref{eq:poten-in-p}. We have 
 \begin{align}
	Y_{\rm{ex}}&=\frac{1}{(2\pi)^3}\int \frac{1}{\bm q^2+\mu_{\rm{ex}}} \left (\frac{\tilde{\Lambda}^2-\mu^2_{\rm{ex}}}{\bm q^2+\tilde{\Lambda}^2}\right )^2 e^{i\bm q\cdot \bm r}d^3\bm q,\notag\\
	&=\frac{1}{4\pi r}({\rm{e}}^{-\mu_{\rm{ex}}r}-{\rm{e}}^{-\tilde{\Lambda}r})-\frac{\tilde{\Lambda}^2-\mu_{\rm{ex}}^2}{8\pi\tilde{\Lambda}}{\rm{e}}^{-\tilde{\Lambda}r},\\
	\bm \sigma\cdot \bm LO_{\rm{ex}}&=\frac{1}{(2\pi)^3}\int\frac{i \bm \sigma\cdot (\bm q\times\bm Q)  }{\bm q^2+\mu^2_{\rm{ex}}}\left (\frac{\tilde{\Lambda}^2-\mu^2_{\rm{ex}}}{\bm q^2+\tilde{\Lambda}^2}\right )^2 e^{i\bm q\cdot \bm r}d^3\bm q\notag\\
	&=\bm \sigma\cdot \bm L\frac{1}{r}\frac{\partial }{\partial r} Y_{\rm{ex}}\label{eq:Fourier-O},
	 \end{align}      
where we use the definition of the angular momentum operator $\bm L$, such as $\bm L=\bm r\times \bm Q$ ~\cite{Liu:2011xc}. The Fourier transformations of the functions $i \bm \epsilon_4^*\cdot (\bm q\times\bm Q)/ (\bm q^2+\mu^2_{\rm{ex}})$ and $i \mathcal{T}\cdot (\bm q\times\bm Q)/ (\bm q^2+\mu^2_{\rm{ex}})$ can be preformed in the similar way as  Eq.~\eqref{eq:Fourier-O} after replacing $\bm \sigma$ with $\bm \epsilon_4^*$ and $ \mathcal{T}$, respectively. Before preforming the Fourier transformation on $\bm \sigma\cdot \bm q \bm \epsilon_4^*\cdot \bm q/(\bm q^2+\mu_{\rm{ex}}^2)$, we can decompose it as 
\begin{align}
	\frac{\bm \sigma\cdot \bm q \bm \epsilon_4^*\cdot \bm q}{\bm q^2+\mu_{\rm{ex}}^2}=\frac{1}{3}\left[\bm \sigma\cdot\bm \epsilon_4^*\left (1-\frac{\mu_{\rm{ex}}^2}{\bm q^2+\mu_{\rm{ex}}^2}\right )-S(\bm \sigma,\bm \epsilon_4^*,\hat q)\frac{|\bm q|^2}{\bm q^2+\mu_{\rm{ex}}^2}\right], \label{eq:CT-pot}
\end{align}
where $S(\bm \sigma,\bm \epsilon_4^*,\hat q)=3 \bm \sigma\cdot \hat q \bm \epsilon_4^*\cdot \hat q-\bm \sigma\cdot\bm \epsilon_4^*$ is the tensor operator in the momentum space. It can be found that the constant term in Eq.~\eqref{eq:CT-pot} leads to the $\delta(\bm r)$ term in coordinate space after the Fourier transformation without form factor. With the form factor, the $\delta(\bm r)$ term can be replaced with the Fourier transformation of the form factor, and it dominates the short-range part of the potential. As a result, the short-range part is heavily depending on the cutoff $\Lambda$~\cite{Thomas:2008ja,Tornqvist:1993ng,Yalikun:2021bfm}. There are several treatments of the $\delta(\bm r)$ in the literature focused on the molecular pentaquarks , $P_c(4312)$, $P_c(4380)$, $P_c(4440)$and $P_c(4457)$. The $\delta(\bm r)$ is fully included in the OBE model in Refs.~\cite{Chen:2019asm,He:2019ify}, and several cutoff parameters are used to reproduce the four $P_c$ pentaquarks. And in Ref.~\cite{Liu:2019zvb}, the $\delta(\bm r)$ is dropped, and the four $P_c$ pentaquarks are reproduced with the same cutoff parameter, but larger values for the coupling constants are used. In Ref.~\cite{Yalikun:2021bfm}, the four $P_c$ pentaquarks are simultaneously reproduced with the same cutoff parameter by introducing a reduction parameter $a$, which adjusts the strength of the short-range part of the potential dominated by the $\delta(\bm r)$ term. In the effective field theory, the short-range contribution cannot be fully captured by the OBE model, which may be viewed as there can be contributions from exchanging heavier particles. 
 The introducing $a$ is an extra subtraction of the regularized potentials. It is equivalent to introducing an extra contact interaction to take into account extra short-range interaction from the other heavier meson exchange. It is introduced as                  
\begin{align}
	\frac{\bm \sigma\cdot \bm q \bm \epsilon_4^*\cdot \bm q}{\bm q^2+\mu_{\rm{ex}}^2}=\frac{1}{3}[&\bm \sigma\cdot\bm \epsilon_4^*\left (1-a-\frac{\mu_{\rm{ex}}^2}{\bm q^2+\mu_{\rm{ex}}^2}\right )\notag\\
	&-S(\bm \sigma,\bm \epsilon_4^*,\hat q)\frac{|\bm q|^2}{\bm q^2+\mu_{\rm{ex}}^2}]. \label{eq:CT-pot1}
\end{align}
After preforming the Fourier transformation on Eq.~\eqref{eq:CT-pot1}, we have~\cite{Yalikun:2021bfm,Wang:2020dya}
\begin{align}
	&\frac{1}{(2\pi)^3}\int \frac{\bm \sigma\cdot \bm q \bm \epsilon_4^*\cdot \bm q}{\bm q^2+\mu_{\rm{ex}}^2} \left (\frac{\tilde{\Lambda}^2-\mu^2_{\rm{ex}}}{\bm q^2+\tilde{\Lambda}^2}\right )^2 e^{i\bm q\cdot \bm r}d^3\bm q\notag\\
	&=-\frac{1}{3}[\bm\sigma\cdot \bm\epsilon_4^*C_{\rm{ex}}+S(\bm\sigma,\bm\epsilon_4^*,\hat r)T_{\rm{ex}}]\label{eq:CT-pot2},
\end{align}
where $S(\bm \sigma,\bm \epsilon_4^*,\hat r)=3 \bm \sigma\cdot \hat r \bm \epsilon_4^*\cdot \hat r-\bm \sigma\cdot\bm \epsilon_4^*$ is the tensor operator in the coordinate space. The functions $C_{\rm{ex}}$ and $T_{\rm{ex}}$ can be expressed as 
\begin{align}
	C_{\rm{ex}}&=\frac{1}{r^2}\frac{\partial}{\partial r}r^2\frac{\partial}{\partial r}Y_{\rm{ex}}+\frac{a}{(2\pi)^3}\int \left (\frac{\tilde{\Lambda}^2-\mu^2_{\rm{ex}}}{\bm q^2+\tilde{\Lambda}^2}\right )^2 e^{i\bm q\cdot \bm r}d^3\bm q,\\
	T_{\rm{ex}}&=r\frac{\partial}{\partial r}\frac{1}{r}\frac{\partial}{\partial r}Y_{\rm{ex}},
	 \end{align}      
where the term proportional to the parameter $a$ can adjust the contribution from the $\delta(\bm r)$ term. $a=0(1)$ means that the contribution of the $\delta(\bm r)$ term is fully included (excluded). The Fourier transformation of the function $\bm \sigma\cdot \bm q  \mathcal{T}\cdot \bm q/(\bm q^2+\mu_{\rm{ex}}^2)$ can be taken in a similar way as  Eq.~\eqref{eq:CT-pot2}. 

With the functions $Y_{\rm{ex}}$, $O_{\rm{ex}}$, $C_{\rm{ex}}$ and $T_{\rm{ex}}$, the potentials in Eq.~\eqref{eq:poten-in-p} can be written in the coordinate space as  
 \begin{widetext}
\begin{subequations}\label{eq:poten}
\begin{itemize}
	\item Type I: $\tilde{\mathcal{P}}\mathcal{B}\to \tilde{\mathcal{P}}\mathcal{B}$
	\begin{align}
		\mathcal{V}_\sigma^{I}&=-\tau_\sigma g_{\mathcal{B}\mathcal{B}\sigma}g_S  [Y_\sigma-\frac{1}{4m_{\mathcal{B}}^2}\bm\sigma\cdot\bm L O_\sigma] ,\label{v11sigma}\\
\mathcal{V}_{V}^{I}&=-\tau_{V}\frac{g_{\mathcal{B}\mathcal{B}{V}}\beta g_V}{ 2}  [Y_{V}+\frac{1+2\kappa_{\mathcal{B}\mathcal{B} V }}{4m_{\mathcal{B}}^2}\bm\sigma\cdot\bm L O_{V}] ,\label{v11vector}
	\end{align}
	\item Type II: $\tilde{\mathcal{P}}\mathcal{B}\to \tilde{\mathcal{P}^*}\mathcal{B}$
	\begin{align}
		\mathcal{V}_{P}^{II}&=-\tau_P\frac{gg_{\mathcal{B}\mathcal{B}P}}{3\sqrt 2 f_\pi m_{P}} [\bm\sigma\cdot \bm\epsilon_4^*C_{P}+S(\bm\sigma,\bm\epsilon_4^*,\hat r)T_{P}] ,\\
		\mathcal{V}_V^{II}&=-\tau_V\frac{g_{\mathcal{B}\mathcal{B} V}\lambda g_V}{2m_{\mathcal{B}}} \left\{\frac{(1+\kappa_{\mathcal{B}\mathcal{B} V})}{3}[2\bm\sigma\cdot \bm\epsilon_4^*C_V
		-S(\bm\sigma,\bm\epsilon_4^*,\hat r)T_V]-(2+3\kappa_{\mathcal{B}\mathcal{B} V })\bm\epsilon_4^*\cdot\bm L O_V\right\} ,
	\end{align}
	\item Type III: $\tilde{\mathcal{P}}^*\mathcal{B}\to \tilde{\mathcal{P}}^*\mathcal{B}$
	\begin{align}
		\mathcal{V}_\sigma^{III}&= -\tau_\sigma g_{\mathcal{B}\mathcal{B}\sigma}g_S\bm{\epsilon}^*_4 \cdot \bm{\epsilon}_2 [Y_\sigma-\frac{1}{4m_{\mathcal{B}}^2}\bm\sigma\cdot\bm L O_\sigma] ,\label{v22sigma}\\
		\mathcal{V}_{P}^{III}&=-\tau_{P}\frac{\sqrt 2gg_{\mathcal{B}\mathcal{B}P}}{6 f_\pi m_{P}} [\bm\sigma\cdot \mathcal{T}C_{P}+S(\bm\sigma,\mathcal T,\hat r)T_{P}] ,\\
		\mathcal{V}_{V}^{III} &=-\tau_{V}\frac{g_{\mathcal{B}\mathcal{B} V}\beta g_V}{ 2}\bm{\epsilon}^*_4 \cdot \bm{\epsilon}_2 [Y_{V}+\frac{1+2\kappa_{\mathcal{B}\mathcal{B}V}}{4m_{\mathcal{B}}^2}\bm\sigma\cdot\bm L O_{V}] \notag\\
		&-\tau_{V}\frac{g_{\mathcal{B}\mathcal{B} V}\lambda g_V }{ 2m_{\mathcal{B}}} \left\{\frac{(1+\kappa_{\mathcal{B}\mathcal{B}V})}{3}[2\bm\sigma\cdot \mathcal{T}C_{V}
		-S(\bm\sigma,\mathcal T,\hat r)T_{V}]+(2-\kappa_{\mathcal{B}\mathcal{B} V })\mathcal{T}\cdot\bm L O_{V}\right\} .\label{v22vector}	\end{align}
\end{itemize}
\end{subequations}
\end{widetext}
The OBE potential matrix for the coupled-channel system, $ D_s^-N-\bar D\Lambda-\bar D\Sigma -\bar D^*\Lambda-\bar D^*\Sigma$, can be constructed with the potentials derived in Eq.~\eqref{eq:poten} and the information given in Table~\ref{tab:coupling}. It is convenient to label the five channels, i.e., $ D_s^-N$, $\bar D\Lambda$, $\bar D\Sigma$, $\bar D^*\Lambda$, $\bar D^*\Sigma$, as the first, second, third, fourth, and fifth channels, respectively. They are sorted simply by their thresholds. For the transition from the $j$th to $k$th channel, the OBE potential can be obtained by summing up all possible light meson exchange potentials, such that   
\begin{align}
\mathcal{V}^{jk}=\mathcal{V}^{jk}_\sigma+\mathcal{V}^{jk}_\pi+\mathcal{V}^{jk}_\eta+\mathcal{V}^{jk}_\rho+\mathcal{V}^{jk}_\omega+\mathcal{V}^{jk}_{\bar K}+\mathcal{V}^{jk}_{\bar K^*},\label{eq:cc-pot-mat}
\end{align}  
where $\mathcal{V}_{\rm{ex}}^{jk}$ refers to the potential for the transition $j\to k$ when the meson exchanged is being the one at the lower index. The potential $\mathcal{V}_{\rm{ex}}^{jk}$ can be obtained from Eq.~\eqref{eq:poten} by replacing the corresponding coupling constants and isospin factors given in Table~\ref{tab:coupling}. For example, the $\rho$ meson exchange potential for the transition $\bar D^*\Sigma\to \bar D^*\Sigma$, which is denoted by $\mathcal{V}_{\rho}^{55}$ in our notation, is obtained by replacing the coupling constants $g_{\mathcal{B}\mathcal{B}V}$ and $\kappa_{\mathcal{B}\mathcal{B}V}$ in Eq.~\eqref{v22vector} with $2g_{NN\rho}\alpha_V$ and $\frac{\kappa_{NN\rho}}{4\alpha_V}$, respectively. For the isospin factor, $\tau_V=-2$ is taken for the potential with $I=1/2$. 

For the masses of the exchanged mesons, we take isospin average masses, which are $m_\pi=137.2$, $m_\eta=547.9$, $m_\rho=775.3$, $m_\omega=782.7$, $m_{\bar K}=493.7$ and $m_{\bar K^*}=891.7$ in the unit of MeV~\cite{ParticleDataGroup:2020ssz}. 
In Ref.~\cite{ParticleDataGroup:2020ssz}, the lightest scalar meson is labeled with $f_0(500)$, which is a broad state and its mass has not been accurately given. In the present work, we simply take $600$ MeV for the $\sigma$ meson mass, and the different choices of its mass from $400$ to $800$ MeV affect the result a little, and can be smeared by a small variation of the cutoff. 
 Thresholds (labeled with $W_j$ for the $j$th channel) and partial wave components of those channels with spin-parities $J^P=1/2^-,3/2^-$ are shown in Table~\ref{tab:mass-parwave}. The notation $^{2S+1}L_J$ is used to identify various partial waves, in which $S$, $L$ and $J$ stand for the spin, orbital and total angular momentums, respectively.
\begin{table*}[htbp!]\centering
	\caption{Thresholds of the channels and partial wave components of $J^P$ states.}\label{tab:mass-parwave}
	\begin{ruledtabular}
		\begin{tabular}{c c c c c c}
			Channels&$ D_s^-N$&$\bar D\Lambda$&$\bar D\Sigma$&$\bar D^*\Lambda$&$\bar D^*\Sigma$\\\hline
			$W_j$[MeV]~\cite{ParticleDataGroup:2020ssz}&$2907.3 $&$ 2982.9 $&$ 3060.4 $&$ 3124.2 $&$ 3201.7$\\\hline
			$J^P=1/2^-$&$^2S_{1/2}$&$^2S_{1/2}$&$^2S_{1/2}$&$^2S_{1/2}$, $^4D_{1/2}$&$^2S_{1/2}$, $^4D_{1/2}$\\
			$J^P=3/2^-$&$^2D_{3/2}$&$^2D_{3/2}$&$^2D_{3/2}$&$^4S_{3/2},^2D_{3/2},^4D_{3/2}$&$^4S_{3/2},^2D_{3/2},^4D_{3/2}$\\
		\end{tabular}
	\end{ruledtabular}
	\end{table*}
In the actual calculation, the spin operators in the potentials should be projected out, and this is done by sandwiching the spin operators between the partial waves of the initial and final states. Since, the partial waves of the channels listed in Table~\ref{tab:mass-parwave} are determined by the spin-parity of the individual hadron and nothing to do with flavors, we can refer the channels \{$ D_s^-N,\bar D\Lambda,\bar D\Sigma$\} to $\tilde{\mathcal{P}}\mathcal{B}$, \{$\bar D^* \Lambda,\bar D^*\Sigma$\} to $\tilde{\mathcal{P}}^*\mathcal{B}$. The partial waves of the $\tilde{\mathcal{P}}\mathcal{B}$ and $\tilde{\mathcal{P}}^*\mathcal{B}$ system with spin-parities $J^P=1/2^-,3/2^-$ are 
\begin{itemize}
	\item $J^P=1/2^-(\tilde{\mathcal{P}}\mathcal{B})$: $ |^2S_{1/2}\rangle$,
	\item $J^P=1/2^-(\tilde{\mathcal{P}}^*\mathcal{B})$: $|^2S_{1/2}\rangle,|^{4}D_{1/2}\rangle$,
	\item $J^P=3/2^-(\tilde{\mathcal{P}}\mathcal{B})$: $ |^2D_{3/2}\rangle$,
	\item $J^P=3/2^-(\tilde{\mathcal{P}}^*\mathcal{B})$: $|^4S_{3/2}\rangle,|^{2}D_{3/2}\rangle,|^{4}D_{3/2}\rangle$.
\end{itemize}
The spin operators for the three types of scattering processes in Eq.~\eqref{eq:poten} are listed in the rows labeled with $\mathcal{O}$ of Table ~\ref{tab:PartialWavesProjections}. The partial wave projection of the operator $\mathcal{O}$ can be done by calculating  
\begin{align}\
	_f\langle ^{2s+1}L'_J | \mathcal{O} | ^{2s+1}L_J \rangle_i,
\end{align}
where $_f\langle ^{2s'+1}L'_J |$ and $| ^{2s+1}L_J \rangle_i$ stand for the partial waves for the final and initial states, respectively. The results are calculated with the technics introduced in the appendix of Ref.~\cite{Yalikun:2021bfm}, and collected in Table ~\ref{tab:PartialWavesProjections}. For instance, the partial wave projections of the operators $\mathcal{O}$ for the process $\tilde{\mathcal{P}}\mathcal{B}\to\tilde{\mathcal{P}}^*\mathcal{B}$ with $J^P=1/2^-$ and $J^P=3/2^-$ are obtained by calculating $(\langle ^{2}S_{1/2} | \mathcal{O} | ^{2}S_{1/2} \rangle,\langle ^{2}S_{1/2} | \mathcal{O} | ^{4}D_{1/2} \rangle)$ and $(\langle ^{2}D_{3/2} | \mathcal{O} | ^{4}S_{3/2} \rangle,\langle ^{2}D_{3/2} | \mathcal{O} | ^{2}D_{3/2} \rangle,\langle ^{2}D_{3/2} | \mathcal{O} | ^{4}D_{3/2} \rangle)$, respectively. Similarly, the partial wave projections for the process $\tilde{\mathcal{P}}^*\mathcal{B}\to\tilde{\mathcal{P}}^*\mathcal{B}$ with $J^P=1/2^-$ and $J^P=3/2^-$ are done by calculating
\begin{align}
\begin{pmatrix}\langle ^{2}S_{1/2} | \mathcal{O} | ^{2}S_{1/2} \rangle&\langle ^{2}S_{1/2} | \mathcal{O} | ^{4}D_{1/2} \rangle\\
	\langle ^{4}D_{1/2} | \mathcal{O} | ^{2}S_{1/2} \rangle&\langle ^{4}D_{1/2} | \mathcal{O} | ^{4}D_{1/2} \rangle
\end{pmatrix}
\end{align}  
and  
\begin{align}
	\begin{pmatrix}\langle ^{4}S_{3/2} | \mathcal{O} | ^{4}S_{3/2} \rangle&\langle ^{4}S_{3/2} | \mathcal{O} | ^{2}D_{3/2} \rangle&\langle ^{4}S_{3/2} | \mathcal{O} | ^{4}D_{3/2} \rangle \\
		\langle ^{2}D_{3/2} | \mathcal{O} | ^{4}S_{3/2} \rangle&\langle ^{2}D_{3/2} | \mathcal{O} | ^{2}D_{3/2} \rangle&\langle ^{2}D_{3/2} | \mathcal{O} | ^{4}D_{3/2} \rangle \\
		\langle ^{4}D_{3/2} | \mathcal{O} | ^{4}S_{3/2} \rangle&\langle ^{4}D_{3/2} | \mathcal{O} | ^{2}D_{3/2} \rangle&\langle ^{4}D_{3/2} | \mathcal{O} | ^{4}D_{3/2} \rangle 
		\end{pmatrix},
	\end{align}      
	respectively. Results are given as the matrix form in Table ~\ref{tab:PartialWavesProjections}.
\begin{table*}[ht!]\centering
	\caption{The partial wave projection of the spin operators in the potentials of Type I, II and III in Eq.~\eqref{eq:poten}.}\label{tab:PartialWavesProjections}
	\begin{ruledtabular}
	\begin{tabular}{c|c|ccc}
$i\to f$&$\tilde{\mathcal{P}}\mathcal{B}\to\tilde{\mathcal{P}}\mathcal{B}$&\multicolumn{3}{c}{$\tilde{\mathcal{P}}\mathcal{B}\to\tilde{\mathcal{P}}^*\mathcal{B}$}	\\	
\hline
$\mathcal{O}$&$\bm\sigma\cdot\bm L$&$\bm\epsilon_4^*\cdot\bm L$&$\bm\sigma\cdot\bm\epsilon_4^*$&$S(\bm\sigma,\bm\epsilon_4^*,\hat r)$\\
$J^P=1/2^-$&$0$&$(0,0)$&$(\sqrt 3,0)$&$(0,\sqrt 6)$\\
$J^P=3/2^-$&$-3$&$(0,-\sqrt{3},-\sqrt 3)$&$(0,\sqrt 3,0)$&$(-\sqrt 3,0,\sqrt 3)$
	\end{tabular}\\
	\begin{tabular}{c|ccccc}
		$i\to f$&\multicolumn{5}{c}{$\tilde{\mathcal{P}}^*\mathcal{B}\to\tilde{\mathcal{P}}^*\mathcal{B}$}\\		
		\hline
		$\mathcal{O}$&$\bm\epsilon_2\cdot\bm\epsilon_4^*$&$\bm\epsilon_2\cdot\bm\epsilon_4^*\bm\sigma\cdot\bm L$&$\bm\sigma\cdot\mathcal{T}$&$S(\bm\sigma,\mathcal{T},\hat r)$&$\mathcal{T}\cdot\bm L$\\
		$J^P=1/2^-$&$\left(
				\begin{array}{cc}
				 1 & 0 \\
				 0 & 1 \\
				\end{array}
				\right)$&$\left(
					\begin{array}{cc}
					 0 & 0 \\
					 0 & -3 \\
					\end{array}
					\right)$&$\left(
					\begin{array}{cc}
					 -2 & 0 \\
					 0 & 1 \\
					\end{array}
					\right)$&$\left(
					\begin{array}{cc}
					 0 & \sqrt{2} \\
					 \sqrt{2} & -2 \\
					\end{array}
					\right)$&$\left(
					\begin{array}{cc}
					 0 & 0 \\
					 0 & -3 \\
					\end{array}
					\right)$\\
		$J^P=3/2^-$&$\left(
			\begin{array}{ccc}
			 1 & 0 & 0 \\
			 0 & 1 & 0 \\
			 0 & 0 & 1 \\
			\end{array}
			\right)$&$\left(
				\begin{array}{ccc}
				 0 & 0 & 0 \\
				 0 & 1 & 2 \\
				 0 & 2 & -2 \\
				\end{array}
				\right)$&$\left(
				\begin{array}{ccc}
				 1 & 0 & 0 \\
				 0 & -2 & 0 \\
				 0 & 0 & 1 \\
				\end{array}
				\right)$&$\left(
					\begin{array}{ccc}
					 0 & -1 & 2 \\
					 -1 & 0 & 1 \\
					 2 & 1 & 0 \\
					\end{array}
					\right)$&$\left(
						\begin{array}{ccc}
						 0 & 0 & 0 \\
						 0 & 1 & 2 \\
						 0 & 2 & -2 \\
						\end{array}
						\right)$
			\end{tabular}
\end{ruledtabular}
\end{table*}  

\begin{figure}[ht!]\centering
	\includegraphics[width=0.5\textwidth]{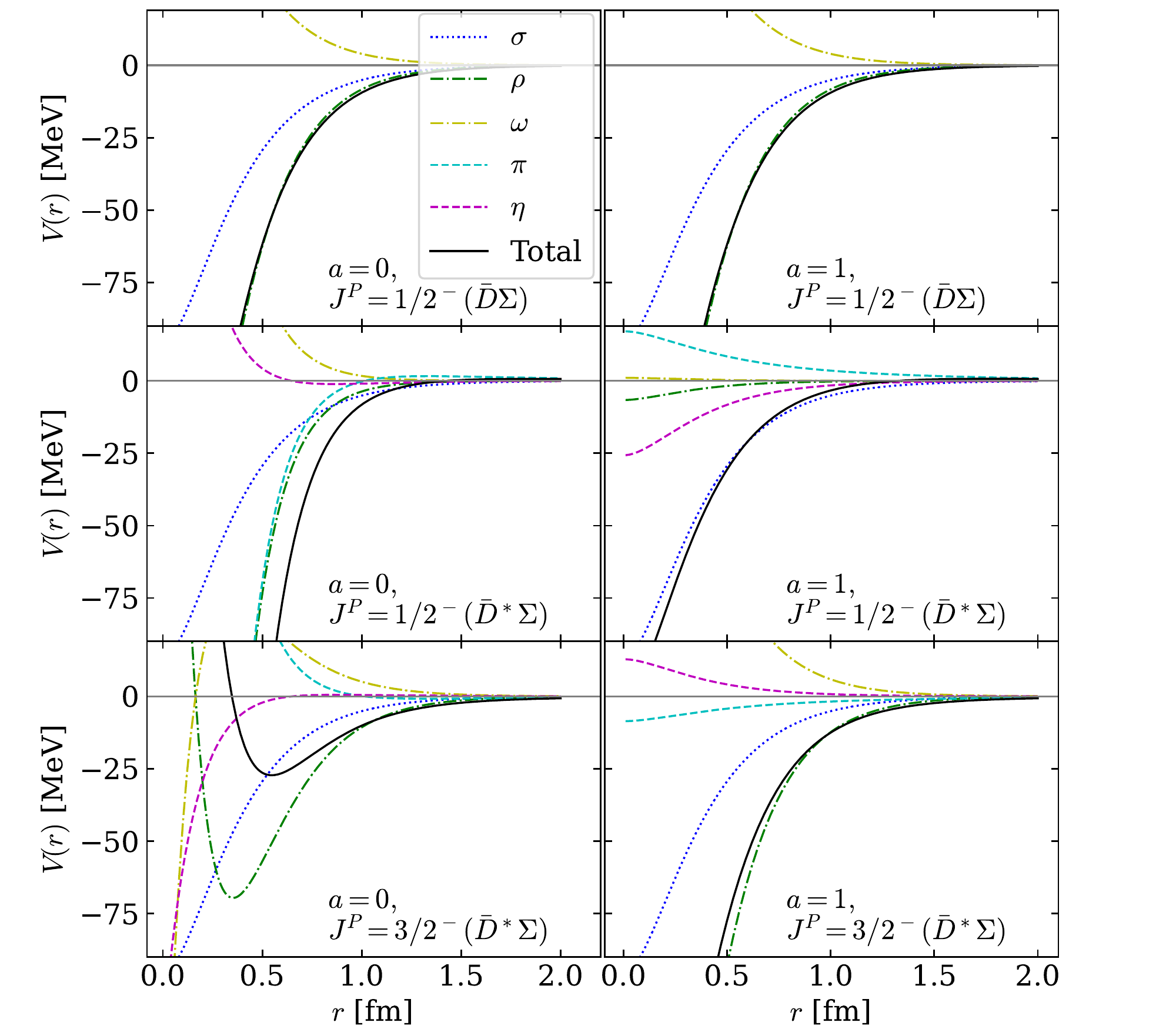}
	\caption{The potentials of $S$-wave states of the $\bar D^{(*)}\Sigma $ system with $I=1/2$  in the function of coordinate $r$, where the cutoff is set to $\Lambda=1.2$ GeV. $a=0$ means the potentials include the full $\delta(\bm r)$ term while $a=1$ means the $\delta(\bm r)$ term is fully removed.}\label{fig_pot}
\end{figure}

To conclude this section, we show the shapes of the potentials of the most important channels, such as $\bar D\Sigma$ and $\bar D^*\Sigma$, under the two extreme treatments of the $\delta(\bm r)$. The $S$ wave potentials of the $\bar D\Sigma$ and $\bar D^*\Sigma$ systems with $I=1/2$ in a function of coordinate $r$ as the cutoff is set to $\Lambda=1.2$ GeV are plotted in  Fig.~\ref{fig_pot}, where the potentials with (without) the $\delta(\bm r)$ term are shown in the left (right) column. 
The potentials for the $\bar D\Sigma$ system with $J^P=1/2^-$ are shown in the first row, while the potentials for the $\bar D^*\Sigma$ system with $J^P=1/2^-$ and $J^P=3/2^-$ are shown at the second and third rows. In each of the subplots, various meson exchange potentials are plotted separately, and the sum of them is also plotted with line labeled ``Total". The potentials of the $\bar D\Sigma$ system are only proportional to the Yukawa term $Y_{\rm{ex}}$ and independent of the $\delta(\bm r)$ term. With the $\delta(\bm r)$ term, the vector and pseudoscalar meson exchange potentials of the $\bar D^*\Sigma$ system with $J^P=1/2^-,3/2^-$ can change their signs once due to the short-range $\delta(\bm r)$ term in their core which has an opposite sign relative to its remaining part. After removing the $\delta(\bm r)$ term, those potentials are consistent in sign in the whole range of $r$. The S wave total potentials in both $\bar D\Lambda$ and $\bar D^*\Lambda$ are repulsive, and there is no bound state accordingly. In addition, the $S$ wave potential for  the $ D_s^-N$ system is attractive due to the $\sigma$ meson exchange alone, but it is not strong enough to form a bound state.   

\section{Results} 
        \label{sec:numerical_results}
In this section, firstly, we describe our procedure of solving the two-body Schr\"odinger equation, and explain the behaviors of the bound states or resonances emerged as poles of the scattering matrix. Secondly, with the OBE potentials of $\bar D\Sigma$ and $\bar D^*\Sigma$ systems derived in the previous section, we investigate the possibilities of the bound states by solving the single channel Schr\"odinger equation. Thirdly, considering the coupled-channel system, $ D_s^-N-\bar D\Lambda-\bar D\Sigma -\bar D^*\Lambda-\bar D^*\Sigma$, we further investigate the resonances below the thresholds of channels $\bar D\Sigma$ and $\bar D^*\Sigma$, and evaluate their partial decay widths. 

For the coupled-channel potential matrix $V_{jk}$, the radial Schr\"odinger equation can be written as  
\begin{eqnarray}
		\left[-\frac{\hbar^2}{2\mu_j}\frac{d^2}{d r^2}+\frac{\hbar^2 l_j(l_j+1)}{2\mu_jr^2} + W_j\right]u_j+
\sum_{k}
	\mathcal{V}_{jk}u_k=
E u_j,
\label{eq_schro_coupl}
\end{eqnarray}
where $j$ is the channel index; $u_j$ is defined by $u_j(r)=rR_j(r)$ with the radial wave function $R_j(r)$ for the $j$th channel; $\mu_j$ and $W_j$ are the corresponding reduced mass and threshold; and$E$ is the total energy of the system. The momentum for channel $j$ is given as 
\begin{align}\label{eq:ch-mom}
	q_j(E)=\sqrt{2\mu_j(E-W_j)}.
\end{align}
 By solving Eq.~\eqref{eq_schro_coupl}, we obtain the wave function which is normalized to satisfy the  incoming boundary condition for the $j$th channel given as~\cite{osti_4661960}
\begin{eqnarray}
	u_{j}^{(k)}(r)\overset{r\rightarrow \infty}{\longrightarrow} \delta_{jk}e^{-iq_j r}-S_{jk}(E)e^{iq_j r}\label{eq:asym-wave},
\end{eqnarray}
where $S_{jk}(E)$ is the scattering matrix component. In the multi-channel problem, there is a sequence of thresholds $W_1<W_2<\cdots$, and the scattering matrix element $S_{jk}(E)$ is an analytic function of $E$ except at the branch points of $E=W_j$ and poles. Bound states and resonances are represented as the poles at $E_{\mathrm{pole}} $ of the $S_{jk}(E)$ in the complex energy plane~\cite{osti_4661960}. 

The characterization of these poles requires to analytically continue the S matrix to the complex energy plane, and the poles should be searched on the correct Riemann sheet. Note that channel momentum $q_j$ is a multivalued function of energy $E$; there are two Riemann sheets in the complex energy plane for each channel; one is called the first or physical sheet, while the other one is called the second or unphysical sheet. In the physical sheet, complex energy $E$ maps to the upper-half plane ($\rm{Im}[q_j]\geq 0$) of the channel momentum $q_j$. In the multi-channel case with different thresholds, bound states can emerge as poles on the real energy axis of the physical sheet at the energy region below the lowest threshold $W_1$. The binding energy labeled as $\mathbb{B}$ in our notation can be evaluated as
\begin{align}
	\mathbb{B}=E_{\mathrm{pole}}-W_1, 
\end{align}
where $E_{\mathrm{pole}}$ is the position of the pole.
In the unphysical sheet, complex energy $E$ maps to the lower-half plane ($\rm{Im}[q_j]< 0$) of the channel momentum $q_j$. Poles may appear in this sheet, and those poles correspond to resonances if their real parts are larger than the thresholds of some channels (these channels are called open channels). Any of the resonance poles has its conjugate pole $E_{\mathrm{pole}}^* $. Among them, the one with a negative imaginary part which is closer to the real energy axis of the physical sheet than the other one has a significant impact on the scattering amplitude (see the review section in ~\cite{ParticleDataGroup:2020ssz}). The real and imaginary
parts of the pole $E_{\mathrm{pole}} $, may be parametrized as the mass and the half-width of the resonance~\cite{Garzon:2012np}, such as
\begin{align}\label{eq:pole}
	E_{\mathrm{pole}}=M-i\Gamma/2. 
\end{align}
Resonance poles that are located on the unphysical sheet closest to the physical sheet are the ones that, together with bound states, are much likely to generate structures in the scattering amplitude. 
\subsection{Bound states} \label{subsec:A}
We start by discussing the bound states in $\bar D\Sigma$ and $\bar D^*\Sigma$ systems within the OBE framework. Bound state energy is obtained by solving the the Schr\"odinger equation in Eq.~\eqref{eq_schro_coupl}, and $S-D$ wave mixing effects are considered. The parameter $a$ is added to adjust the short-range contribution from $\delta(\bm r)$ in the OBE model. On the other hand, the parameter $a$ also plays a similar role as the phenomenological contact term which is used to determine the short-range dynamics of the hadron interaction~\cite{Du:2021fmf}. Bound state energy of the $\bar D\Sigma$ and $\bar D^*\Sigma$ molecular isodoublet systems with spin parity $J^P=1/2^-$ and $J^P=3/2^-$ is evaluated by varying the cutoff $\Lambda$ after several values for $a$ are taken. 

Table~\ref{tab:DbSig_BE_OBE} shows the behavior of the bound state energies within the cutoff range $1-1.8$ GeV, in which the binding energies of corresponding single channel systems (given in the parentheses) are listed at the columns labeled with $\mathbb{B}$. Here, we list the results for four different $\delta(\bm r)$ term contributions, $a=0,0.58,0.78$ and  $a=1$. The $\delta(\bm r)$ term is included in the OBE potentials by setting $a=0$ and it is fully excluded by taking $a=1$. The case with $a=0.58$ or $a=0.78$ corresponds to that $\delta(\bm r)$ term in whole OBE potentials reduced by $58$\% or $78$\%, respectively. The two parameters are constrained in the our previous work in Ref.~\cite{Yalikun:2021bfm} to simultaneously reproduce the masses of the $P_c(4440)$ and $P_c(4457)$ with the same cutoff in the single channel $\bar D^*\Sigma_c$. The binding energy of the $\bar D\Sigma$ bound state with $J^P=1/2^-$ is independent of the $\delta(\bm r)$ term, because the OBE potential of this system is free from the $\delta(\bm r)$ term.
For the two bound states in the $\bar D^*\Sigma$ system with spin parities $J^P=1/2^-$ and $J^p=3/2^-$, the different reduction of the $\delta(\bm r)$ term has a large effect on the binding energy. The binding energy is heavily dependent on the cutoff $\Lambda$ when the $\delta(\bm r)$ term is fully included in the OBE potentials. As the value of $a$ increases, $1/2^-(\bar D^*\Sigma)$ state tends to be a shallow bound state, while the $3/2^-(\bar D^*\Sigma)$ state, by contrast, tends to be deep bound state. The reason is that, for the $\bar D^*\Sigma$ system, the total potential with $J^P=1/2^-$ gets shallower as the parameter $a$ increases, which leads to smaller binding energy, while the situation is reversed for the potential with $J^P=3/2^-$. Furthermore, three bound states, $1/2^-(\bar D\Sigma)$, $1/2^-(\bar D^*\Sigma)$ and $3/2^-(\bar D^*\Sigma)$ are simultaneously bound with the cutoff $1.5$ GeV with either $a=0.58$ or $a=0.78$. In addition, the $S-D$ wave mixing effects on the $J^P=3/2^-(\bar D^*\Sigma )$ state are relatively larger than that on $J^P=1/2^-(\bar D^*\Sigma )$ state. 
 
\begin{table}[htbp!]\centering
	\caption{Binding energy ($\mathbb{B}$) of the bound states in the single channel $\bar D\Sigma$ and $\bar D^*\Sigma$ systems with isospin $I=1/2$ as a function of cutoff $\Lambda$ after fixing the value for $a$. Each entry with a ``$\cdots$'' means that the potentials are not strong  enough to form a bound state. The values of cutoff and binding energy are in units of GeV and MeV, respectively.}\label{tab:DbSig_BE_OBE}
	\begin{ruledtabular}
	\begin{tabular}{c|c|c|cc|cc}
		\multirow{3}{*}{$a$}&\multirow{3}{*}{$\Lambda$}&$\mathbb{B}(\bar D\Sigma)$& \multicolumn{4}{c}{$\mathbb{B}(\bar D^*\Sigma)$} \\
		&&$S$-wave&\multicolumn{2}{c}{$S$-wave}&\multicolumn{2}{c}{$S$-$D$ wave mixing}\\
		&&$J^P=1/2^-$&$1/2^-$&$3/2^-$&$1/2^-$&$3/2^-$\\
		\hline
\multirow{4}{*}{$0.0$}
& $ 1.0  $ & $ \cdots $ & $    -7.35 $ & $ \cdots $ & $    -8.96 $ & $ \cdots  $\\
& $ 1.25 $ & $ \cdots $ & $  -125.5  $ & $ \cdots $ & $  -129.23 $ & $ \cdots  $\\
& $ 1.5  $ & $  -0.66 $ & $ <-500    $ & $ \cdots $ & $ <-500    $ & $ \cdots  $\\
& $ 1.8  $ & $ -11.11 $ & $ <-500    $ & $ \cdots $ & $   \cdots $ & $  -1.22  $\\
\hline
\multirow{4}{*}{$0.58$}
& $ 1.0  $ & $ \cdots $ & $  \cdots  $ & $ \cdots $ & $   \cdots $ & $  \cdots $\\
& $ 1.25 $ & $ \cdots $ & $   -0.11  $ & $ \cdots $ & $    -1.1  $ & $  \cdots $\\
& $ 1.5  $ & $  -0.66 $ & $  -26.3   $ & $  -0.18 $ & $   -32.06 $ & $   -5.02 $\\
& $ 1.8  $ & $ -11.11 $ & $ -130.7   $ & $  -5.57 $ & $  -140.55 $ & $  -19.07 $\\ 
\hline
\multirow{4}{*}{$0.78$}
& $ 1.0  $ & $ \cdots $ & $\cdots    $ & $\cdots  $ & $ \cdots   $ & $ \cdots  $\\
& $ 1.25 $ & $ \cdots $ & $\cdots    $ & $\cdots  $ & $ \cdots   $ & $  -0.68  $\\
& $ 1.5  $ & $  -0.66 $ & $\cdots    $ & $ -3.33  $ & $  -0.05   $ & $ -12.17  $\\
& $ 1.8  $ & $ -11.11 $ & $ -6.37    $ & $-18.31  $ & $ -12.58   $ & $ -37.14  $\\
\hline
\multirow{4}{*}{$1.0$}
& $ 1.0  $ & $ \cdots $ & $ \cdots   $ & $\cdots  $ & $\cdots    $ & $\cdots   $\\
& $ 1.25 $ & $ \cdots $ & $ \cdots   $ & $  -0.2  $ & $\cdots    $ & $ -3.56   $\\
& $ 1.5  $ & $  -0.66 $ & $ \cdots   $ & $ -13.42 $ & $\cdots    $ & $-26.17   $\\
& $ 1.8  $ & $ -11.11 $ & $ \cdots   $ & $ -48.0  $ & $\cdots    $ & $-71.73   $\\
	\end{tabular}
\end{ruledtabular}
\end{table}     

The masses of the $P_c$ states observed in the experiment are close to the thresholds of the $\bar D\Sigma_c$ and $\bar D^*\Sigma_c$ channels. In the OBE framework, these two channels support three bound states which can reproduce the masses of the three $P_c$ states, $P_c(4312)$, $P_c(4440)$ and $P_c(4457)$  with proper values of cutoff~\cite{He:2019rva,Chen:2019asm,Liu:2019zvb,Du:2021fmf,Yalikun:2021bfm}. It indicates that, the existence of the bound states formed by the single channel interaction may be a hint for experimental observation. Compared to the bound states found in the single channel $\bar D\Sigma_c$ and $\bar D^*\Sigma_c$ systems in Ref.~\cite{Yalikun:2021bfm}, the discussed three bound states in the single channel $\bar D\Sigma$ and $\bar D^*\Sigma$ systems have similar formation mechanism, and support the existence of the anticharmed strange pentaquarks proposed in Refs.~\cite{Lipkin:1987sk,Gignoux:1987cn}. In addition, a bound state which is $220~\rm{MeV}$ below the $ D_s^-N$ threshold has been found after considering the coupled-channel dynamics of $ D_s^-N-\bar D\Lambda-\bar D\Sigma$  in Ref.~\cite{Hofmann:2005sw}, and which state is strongly coupled to the $\bar D\Sigma$ channel compared to the other two channels. It indicates that the $\bar D\Sigma$ channel provides a more attractive force than others. Thus, a loose bound state below the $\bar D\Sigma$ threshold, which is different from the one found in Ref.~\cite{Hofmann:2005sw} based on the assumption of the $SU(4)$ symmetry, may be bound first in the single channel $\bar D\Sigma$ interaction. 
                              
\subsection{Resonances}
With the coupled-channel potentials of the system $ D_s^-N-\bar D\Lambda-\bar D\Sigma -\bar D^*\Lambda-\bar D^*\Sigma$ obtained in Eq.~\eqref{eq:cc-pot-mat}, we solve the Schr\"odinger equation in Eq.~\eqref{eq_schro_coupl}, and the energy dependent $S(E)$ matrix is extracted from the asymptotic wave function in Eq.~\eqref{eq:asym-wave} (see Refs.~\cite{Kamiya:2019uiw,osti_4661960}). It is seen that the poles of the $S(E)$ matrix on the physical sheet correspond to bound states. Now, we go to the unphysical sheets to search for the poles by analytic continuation of the $S(E)$ matrix, and then evaluate their partial decay widths. In the coupled-channel system $ D_s^-N-\bar D\Lambda-\bar D\Sigma -\bar D^*\Lambda-\bar D^*\Sigma$, the poles are searched for around the thresholds of $\bar D\Sigma$ and $\bar D^*\Sigma$ channels. Those bound states of these two channels as listed in Table~\ref{tab:DbSig_BE_OBE} turn out to be resonances when subthreshold coupled channels are taken into account. These resonances are easily found near the two thresholds. For the $n$ channels system, there are $2^n$ Riemann sheets in the complex energy plane, which can be defined by the imaginary part of the momentum $q_j(E)$ of the $j$th channel (see chapter 20 of Ref.~\cite{osti_4661960} for more details). In our case of the five channels system, there are $2^5$ Riemann sheets, and we focus on the two of them, one is called the physical sheet of the $\bar D\Sigma$ channel ($\rm{Im}[q_1]<0$, $\rm{Im}[q_2]<0$, $\rm{Im}[q_3]\geq 0$, $\rm{Im}[q_4]\geq 0$, $\rm{Im}[q_5]\geq 0$), and the another one is called the physical sheet of the $\bar D^*\Sigma$ channel ($\rm{Im}[q_1]<0$, $\rm{Im}[q_2]<0$, $\rm{Im}[q_3]< 0$, $\rm{Im}[q_4]< 0$, $\rm{Im}[q_5]\geq 0$). These two sheets are also close to the real energy axis of the physical sheet (imaginary parts of the momenta for all channels are positive). The poles below the $\bar D\Sigma$ threshold are searched in the physical sheet of the $\bar D\Sigma$ channel, while the poles below the $\bar D^*\Sigma$ threshold are searched in the physical sheet of the $\bar D^*\Sigma$ channel. 

To evaluate the partial widths of the poles decaying to open channels, first, we calculate the residues of the poles of amplitude $T(E)$. The $S(E)$ matrix has the relation with $T(E)$ ~\cite{Doring:2009yv,Garzon:2012np},
\begin{align}
	S_{jk}(E)=1+i\sqrt{2\rho_j}T_{jk}(E)\sqrt{2\rho_k},
\end{align} 
where $j$ and $k$ are channel indices. In nonrelativistic approximation, two body phase space factor $\rho_j$ for channel $j$ can be written as a function of channel momentum $q_j(E)$ in Eq.~\eqref{eq:ch-mom} as
\begin{align}
	\rho_j=\frac{q_j(E)}{8\pi E}.
\end{align}
The residue matrix $R_{jk}$ of the pole $E_{\mathrm{pole}}$ can be extracted as 
\begin{align}
	R_{jk}=	\lim\limits_{E\to E_{\mathrm{pole}}} (E^2-E^2_{\mathrm{pole}})T_{jk}(E)=g_jg_k,
\end{align}
where $g_j$ is a pole coupling of the $j$th channel. The partial decay widths of the open channels can be calculated as~\cite{Sakai:2019qph,Garzon:2012np}
\begin{align}
	\Gamma_j=\frac{q_j(M)}{8\pi M^2}|g_j|^2,\label{eq:partila-decay}
\end{align}
where $M$ is the real part of the pole.

The positions of the poles and partial decay widths as the function of cutoff $\Lambda$ are shown in Tables~\ref{tab:res-a-0.58} and ~\ref{tab:res-a-0.78}, in which we fix the values of the parameter $a$ to $0.58$ for Table~\ref{tab:res-a-0.58} and $0.78$ for Table~\ref{tab:res-a-0.78}. 
These two values for the parameter $a$ are also used in Ref.~\cite{Yalikun:2021bfm} to simultaneously reproducing the $P_c(4400)$ and $P_c(4457)$ pentaquark masses. Among these poles, the first pole with $J^P=1/2^-$ below the $\bar D\Sigma$ threshold is found in the physical sheet of $\bar D\Sigma$ channel, while the second and third poles with spin parity $J^P=1/2^- $ and $J^P=3/2^-$ are found in the physical sheet of $\bar D^*\Sigma $ channel. The pole positions and partial decay widths of the first, second, and third poles are given in the multicolumns labeled as $J^P=1/2^-(\bar D\Sigma)$, $J^P=1/2^-(\bar D^*\Sigma)$, and $J^P=3/2^-(\bar D^*\Sigma)$, respectively. The dominant channel which couples with poles stronger than other channels, is given in the parenthesis. In the case with $a=0.58$, with cutoff $\Lambda=1.2$ GeV, the first pole is located at $3060.3-i0.3$, while the second and third poles are located at $3197.3-i3.9$ and $3201.2-i2.8$ respectively. For the first pole, there are two S-wave open channels, which are $ D_s^-N$ and $\bar D\Lambda$ channels. Among these channels, the first pole prefers to decay into the $ D_s^-N$ channel as shown in Table~\ref{tab:res-a-0.58}. For the second pole, five channels (four S-wave channels $ D_s^-N$, $\bar D\Lambda$, $\bar D\Sigma$ and $\bar D^*\Lambda$, and one D-wave channel $\bar D^*\Lambda$) are opened. For the third pole, there are six open channels ($ D_s^-N$, $\bar D\Lambda$, $\bar D\Sigma$ in D-wave, one S-wave and two D-wave $\bar D^*\Lambda$). It is also seen from the results in Table~\ref{tab:res-a-0.58} that the S-wave $\bar D\Sigma$ channel is a dominant decay channel for both the second and third poles, and the dominant decay channel remains the same as the cutoff increases except that it changes from a $\bar D\Sigma$ to a $\bar D^*\Lambda$ channel in the S-wave for the third pole when cutoff is up to $1.4$ GeV. 

The three poles have a similar behavior that they tend to move away from the thresholds of their dominant channels in the complex energy plane as the cutoff increases. In other words, the masses of the poles decrease and their half widths increase as the cutoff increases. The reason is that, these poles are associated with the bound states given in Table~\ref{tab:DbSig_BE_OBE} which is inferred from their dominant channels, and the pole masses ($M$) behave similar as the bound state masses ($\mathbb{B}+W$, where $W$ is the threshold of the channel for the bound state in the single channel interaction discussed in Sec.~\ref{subsec:A}). The partial widths of the poles decaying to open channels, which are calculated with Eq.~\eqref{eq:partila-decay}, imply that the half widths of the poles are proportional to the pole couplings $g_j$. For the results presented in Table~\ref{tab:DbSig_BE_OBE}, the magnitude of the pole couplings for the open channels which provide large contributions to their widths, increases together with the cutoff, so the half widths also increase. At the energy region much above the thresholds of open channels, the impact of the phase space factor is not significant. A similar phenomenon can be seen in Ref.~\cite{Yu:2018yxl}, which is governed by the complicated structure of the coupled-channel potential matrix. 

The results corresponding to $a=0.78$ are shown in Table~\ref{tab:res-a-0.78}. This value for $a$ is also taken in Ref.~\cite{Yalikun:2021bfm} to simultaneously reproduce the mass spectra of the three observed $P_c$ states~\cite{LHCb:2019kea}, $P_c(4312)$, $P_c(4440)$ and $P_c(4457)$ with the same cutoff, and it is mentioned that larger value for $a$ is favorable after their widths are taken into account. It is seen from the results in Table~\ref{tab:res-a-0.78} that, the first pole is not sensitive to the parameter $a$ compared to the results in Table~\ref{tab:res-a-0.58} due to the independence behavior of the dominant channel potential on $a$ shown in the first column of Fig.~\ref{fig_pot}, and the minor changes can be understood as the coupled-channel effect. For second and third poles, the mass ordering  is reversed compared to the results in Table~\ref{tab:res-a-0.58} due to the similar mechanism explained in Sec.~\ref{subsec:A}. A similar phenomenon can be found in Ref.~\cite{Yalikun:2021bfm} that, among the two poles near $\Sigma_c\bar D^*$ threshold, the pole with $J^P=1/2$ is higher than the poles with $J^P=3/2^-$ for large $a$ while the situation is reversed with small $a$. Besides, the results in Table~\ref{tab:res-a-0.78} also indicate that the $S$ wave $\bar D^*\Lambda$ channel is the dominant decay channel for the second and third poles while the $ D_s^-N$ channel is the dominant decay channel for the first pole. 

Basically, in our calculation, we can determine neither the cutoff $\Lambda$ nor the reduction parameter $a$, because there is no experimental data for anticharmed strange pentaquarks. But the cutoff ranges taken in our work are somehow reasonable, as the LHCb $P_c$ pentaquakrs\cite{LHCb:2019kea} are reproduced with $\Lambda=1.4$ GeV in Ref.~\cite{Yalikun:2021bfm}, with $\Lambda=1.04$ and $\Lambda=1.32$ in Ref.~\cite{Chen:2019asm}. In Ref.~\cite{Tornqvist:1993ng}, it is mentioned that in nucleon-nucleon interactions values for $\Lambda$ between 0.8 and 1.5 GeV have been
used depending on the model and application, and the larger values ($\Lambda > 1.5$ GeV) are also required for nucleon-nucleon phase shifts. For the parameter $a$, we simply follow the suggestion in Ref.~\cite{Yalikun:2021bfm}, and it is also noted that small variation of the parameter $a$ can change the results presented in the present work by a few percent. Besides, the $ D_s^-N$ channel is the lowest channel which the anticharmed strange pentaquarks can strongly decay to. The first pole dominantly decays to the $ D_s^-N$ channel, and it implies that the production rate of this channel is larger than the other channel, and may be easily detected. For the second and third poles, partial decay widths of the $ D_s^-N$ channel are tiny, but this channel stands out as a sharp peak and can be easily distinguished from the background signal in the experiment with high luminosity.                             

\begin{table*}[htbp!]
	\caption{Pole positions ($M-i\Gamma/2$) and partial decay widths $\Gamma_i$ for each of the open channels in the isodoublet system with spin parity $J^P$ by varying cutoff $\Lambda$ when $a=0.58$ is taken. The channel given in the parentheses correspond to the dominant channel. The partial decay widths for  the partial wave channels in the same hadron pair are given separately. From left to right in each multicolumn $\Gamma_i$, the partial decay widths correspond to the channels \{$ D_s^-N(^2S_{1/2}),\bar D\Lambda(^2S_{1/2})$\} for the first pole $J^P=1/2^-(\bar D\Sigma)$, \{$ D_s^-N(^2S_{1/2}),\bar D\Lambda(^2S_{1/2}),\bar D\Sigma(^2S_{1/2}),\bar D^*\Lambda(^2S_{1/2}),\bar D^*\Lambda(^4D_{1/2})$\} for the second pole $J^P=1/2^-(\bar D^*\Sigma)$ and \{$ D_s^-N(^4D_{3/2}),\bar D\Lambda(^4D_{3/2}),\bar D\Sigma(^4D_{3/2}),\bar D^*\Lambda(^4S_{3/2}),\bar D^*\Lambda(^2D_{3/2}),\bar D^*\Lambda(^4D_{3/2})$\} for the third pole $J^P=3/2^-(\bar D^*\Sigma)$. Mass and width are in units of MeV.   \label{tab:res-a-0.58} }
	\begin{ruledtabular}
	\begin{tabular}{c|c|cc|c|ccccc|c|ccccccc}
		\multirow{2}{*}{$\Lambda$[GeV]}& \multicolumn{3}{c|}{$J^P=1/2^-(\bar D\Sigma)$}& \multicolumn{6}{c|}{$J^P=1/2^-(\bar D^*\Sigma)$} & \multicolumn{7}{c}{$J^P=3/2^-(\bar D^*\Sigma)$}\\
		& $M-i\Gamma/2$ & \multicolumn{2}{c|}{$\Gamma_i$} & $M-i\Gamma/2$ & \multicolumn{5}{c|}{$\Gamma_i$} &  $M-i\Gamma/2$ & \multicolumn{6}{c}{$\Gamma_i$} \\
		\hline
		$ 1.2  $ & $ 3060.3-i0.3  $ & $   0.6 $ & $  0.1 $ & $  3197.3-i3.9  $ & $  0.3 $ & $  0.7 $ & $   6.1 $ & $  0.7 $ & $  0.6 $ & $  3201.2-i2.8 $ & $  0.0 $ & $  1.9  $ & $ 2.3 $ & $  1.2  $ & $ 0.2  $ & $ 1.4 $\\
		$ 1.25 $ & $ 3059.0-i1.2  $ & $   2.3 $ & $  0.3 $ & $  3190.2-i7.1  $ & $  0.7 $ & $  1.1 $ & $  11.3 $ & $  1.2 $ & $  0.5 $ & $  3199.3-i3.7 $ & $  0.0 $ & $  2.2  $ & $ 2.7 $ & $  1.7  $ & $ 0.2  $ & $ 1.5 $\\
		$ 1.3  $ & $ 3055.8-i3.0  $ & $   5.7 $ & $  0.7 $ & $  3179.0-i12.1 $ & $  1.5 $ & $  1.8 $ & $  18.8 $ & $  2.1 $ & $  0.3 $ & $  3196.9-i4.5 $ & $  0.0 $ & $  2.5  $ & $ 3.0 $ & $  2.3  $ & $ 0.2  $ & $ 1.5 $\\
		$ 1.35 $ & $ 3049.6-i6.2  $ & $  11.5 $ & $  1.4 $ & $  3162.0-i20.1 $ & $  3.5 $ & $  3.1 $ & $  29.1 $ & $  3.5 $ & $  0.1 $ & $  3194.0-i5.1 $ & $  0.0 $ & $  2.7  $ & $ 3.2 $ & $  2.9  $ & $ 0.2  $ & $ 1.5 $\\
		$ 1.4  $ & $ 3036.8-i10.6 $ & $  18.9 $ & $  2.5 $ & $  3136.1-i32.1 $ & $  7.8 $ & $  5.5 $ & $  38.7 $ & $  3.8 $ & $  0.2 $ & $  3190.6-i5.6 $ & $  0.0 $ & $  2.9  $ & $ 3.4 $ & $  3.5  $ & $ 0.1  $ & $ 1.4 $\\	
	\end{tabular}
\end{ruledtabular}
\end{table*}

\begin{table*}[htbp!]
	\caption{Similar table as Table~\ref{tab:res-a-0.58} while $a=0.78$ is taken. Each entry with a ``$\cdots$'' means that the pole goes to other Riemann sheet far away from the physical sheet.   \label{tab:res-a-0.78} }
	\begin{ruledtabular}
	\begin{tabular}{c|c|cc|c|ccccc|c|ccccccc}
		\multirow{2}{*}{$\Lambda$[GeV]}& \multicolumn{3}{c|}{$J^P=1/2^-(\bar D\Sigma)$}& \multicolumn{6}{c|}{$J^P=1/2^-(\bar D^*\Sigma)$} & \multicolumn{7}{c}{$J^P=3/2^-(\bar D^*\Sigma)$}\\
		& $M-i\Gamma/2$ & \multicolumn{2}{c|}{$\Gamma_i$} & $M-i\Gamma/2$ & \multicolumn{5}{c|}{$\Gamma_i$} &  $M-i\Gamma/2$ & \multicolumn{6}{c}{$\Gamma_i$} \\
		\hline
		$ 1.2  $ & $  3060.4-i0.1  $ & $  0.2 $ & $ 0.0 $ & $    \cdots     $ & $\cdots$ & $\cdots$ & $\cdots$ & $\cdots $ & $\cdots$ & $  3199.0-i4.9  $ & $ 0.0 $ & $  2.4 $ & $  2.6 $ & $   4.4 $ & $  0.2 $ & $  1.6 $\\
		$ 1.3  $ & $  3058.5-i1.3  $ & $  2.4 $ & $ 0.3 $ & $    \cdots     $ & $\cdots$ & $\cdots$ & $\cdots$ & $\cdots $ & $\cdots$ & $  3191.8-i7.5  $ & $ 0.0 $ & $  2.9 $ & $  3.0 $ & $   7.9 $ & $  0.2 $ & $  1.5 $\\
		$ 1.4  $ & $  3053.9-i4.2  $ & $  7.8 $ & $ 1.0 $ & $    \cdots     $ & $\cdots$ & $\cdots$ & $\cdots$ & $\cdots $ & $\cdots$ & $  3181.5-i9.6  $ & $ 0.0 $ & $  3.2 $ & $  3.2 $ & $  11.6 $ & $  0.1 $ & $  1.2 $\\
		$ 1.55 $ & $  3044.6-i13.4 $ & $ 23.7 $ & $ 4.2 $ & $ 3200.6-i21.1 $ & $  0.3 $ & $ 23.6 $ & $  1.6 $ & $  30.1 $ & $  0.6 $ & $  3160.9-i11.2 $ & $ 0.1 $ & $  3.1 $ & $  3.2 $ & $  14.9 $ & $  0.0 $ & $  0.5 $\\
		$ 1.6  $ & $  3041.9-i17.7 $ & $ 30.6 $ & $ 6.2 $ & $ 3193.6-i26.5 $ & $  0.1 $ & $ 28.9 $ & $  2.0 $ & $  32.2 $ & $  0.5 $ & $  3152.5-i11.2 $ & $ 0.1 $ & $  3.0 $ & $  3.4 $ & $  14.7 $ & $  0.0 $ & $  0.3 $\\	\end{tabular}
	\end{ruledtabular}
\end{table*}
\section{Conclusion}   
Stimulated by the experiment evidence of the LHCb hidden-charm pentaquarks, we investigate the molecular structure of the $P_{\bar cs}$ pentaquarks from the OBE model. The potentials for the systems of $\bar D^{(*)}\Sigma$, $\bar D^{(*)}\Lambda$ and $ D_s^-N$ are constructed with an effective Lagrangian taking into account of HQSS,  SU(3)  flavor symmetry, and all possible light meson exchange dynamics. The dipole form factor as a function of the phenomenological parameter $\Lambda$ is used to regularize the potentials. The short-range contribution from the $\delta(\bm r)$ term is parametrized with a parameter $a$, and it can mimic the role of the contact term used in effective field theory. The possible bound states in the single channels ($\bar D\Sigma$ and $\bar D^*\Sigma$) are searched for with various cutoff $\Lambda$ and parameter $a$. The resonance parameters associated with those bound states are calculated after taking into account the coupled-channel system $ D_s^-N-\bar D\Lambda-\bar D\Sigma -\bar D^*\Lambda-\bar D^*\Sigma$.    

There are three bound states found in the $\bar D\Sigma$ and $\bar D^*\Sigma$ systems with isospin $I=1/2$. Among them, one is identified with the spin parity $J^P=1/2^-$ below $\bar D\Sigma$ threshold, and the other two are identified with $J^P=\{1/2^-,3/2^-\}$ below the $\bar D^*\Sigma$ threshold. The $J^P=1/2^-$ bound state below the $\bar D\Sigma$ threshold can be bound when the cutoff $\Lambda$ is above $1.5$ GeV, and its binding energy is independent of the parameter $a$. The binding energies of the two bound states, $J^P=1/2^-$ and $J^P=3/2^-$ below $\bar D^*\Sigma$ threshold, depend on the parameter $a$, because the potentials of the $\bar D^*\Sigma$ system have contribution of the $\delta(\bm r)$ term. In this system, when the $\delta(\bm r)$ term is fully kept with $a=0$, the binding energy of the $J^P=1/2^-$ state is heavily depending on the cutoff $\Lambda$ while the state with $J^P=3/2^-$ cannot be bound until the cutoff increases up to $1.8$ GeV. As the value of $a$ increases, the  $J^P=1/2^-$ bound state tends to be shallower and the $J^P=3/2^-$ bound state tends to be deeper. It is caused by the sign difference of the $\delta(\bm r)$ term between the potentials corresponding to the $J^P=1/2^-$ and $J^P=3/2^-$ systems. Until now, there has been no experimental data for any $P_{\bar cs}$ pentaquarks, therefore, we simply take $a=0.58$ and $a=0.78$ by following the argument in Ref.~\cite{Yalikun:2021bfm}, in which the masses of the $P_c(4440)$ and $P_c(4457)$ pentaquarks are simultaneously reproduced with the same cutoff in the $\bar D^*\Sigma$ single channel system. As a result, the three states above begin to be bound with cutoff $\Lambda=1.5$ GeV.    

The decay widths of the resonances associated with those bound states are evaluated considering the coupled-channel system $ D_s^-N-\bar D\Lambda-\bar D\Sigma -\bar D^*\Lambda-\bar D^*\Sigma$. With values of $a=0.58$ or $a=0.78$, widths of these resonances emerged as the poles of the $S$ matrix are calculated by varying the cutoff $\Lambda$. The first pole with $J^P=1/2^-$ below $\bar D\Sigma $ threshold decay dominantly to the $ D_s^-N$ channel. It may be easily detected in the process $\bar B_s^0\to \bar n  D_s^- p$. For the second pole with $J^P=1/2^-$ and the third pole with $J^P=3/2^-$ below the $\bar D^*\Sigma$ threshold, partial decay widths of $ D_s^-N$ channel are small. Detecting them in this channel may require much higher statistics. In addition, the mass ordering of the second and third poles is interchanged in the cases with these two values of $a$. The predicted masses and decay widths of those three states may provide valuable information for discovering the $P_{\bar cs}$ pentaquarks in future experiments.

\begin{acknowledgments}
 
	We thank professors Feng-Kun Guo and Jia-Jun Wu for helpful discussions. This work is supported by the NSFC and the Deutsche Forschungsgemeinschaft (DFG, German Research Foundation) through the funds provided to the Sino-German Collaborative Research Center TRR110 Symmetries and the Emergence of Structure in QCD (NSFC Grant No. 12070131001, DFG Project-ID 196253076-TRR 110), by the NSFC Grants No. 11835015 and No. 12047503, and by the Chinese Academy of Sciences (CAS) under Grant No. XDB34030000.
   
   \end{acknowledgments}

   \section*{Appendix}
   \begin{appendices}
	\section{Explicit form of the Octet Lagrangian}\label{appen:lag}

	Following Refs.~\cite{deSwart:1963pdg,Ronchen:2012eg}, we show the explicit form of the Lagrangian in Eqs.~\eqref{lag:BBV} and ~\eqref{lag:BBP} below. For the SU(3) octet baryon and pseudoscalar scalar meson interaction, the effective Lagrangian can be written as  
	\begin{align}
	\mathcal{L}_{\mathcal{B} \mathcal{B} P}&=-\frac{\sqrt{2}D}{m_P}\langle\bar{\mathcal{B}} \gamma^5\gamma_\mu [\partial^\mu\mathbb{P},\mathcal{B}]_+\rangle-\frac{\sqrt{2}F}{m_P}\langle\bar{\mathcal{B}} \gamma^5\gamma_\mu [\partial^\mu\mathbb{P},\mathcal{B}]_-\rangle,\label{appen:eq:pseudoscalar}
	\end{align}
	where the symbol $\langle\cdots\rangle$ denotes trace of SU(3) matrices, and $[A,B]_\pm=AB\pm BA$. SU(3) matrices for octet baryon and pseudoscalar meson are \cite{Pich:1995bw,Bernard:1995dp,ParticleDataGroup:2020ssz}
	\begin{align}
		\mathcal{B}&=\begin{pmatrix}\frac{\Sigma^0}{\sqrt 2}+\frac{\Lambda}{\sqrt 6}&\Sigma^+&p\\\Sigma^-&-\frac{\Sigma^0}{\sqrt 2}+\frac{\Lambda}{\sqrt 6}&n\\\Xi^-&\Xi^0&-\sqrt{\frac{2}{3}}\Lambda\end{pmatrix},\\
		\mathbb{P}&=
		\begin{pmatrix}
			\frac{\pi^0}{\sqrt 2}+\frac{\eta}{\sqrt 6}&\pi^+&K^+\\
			\pi^-&-\frac{\pi^0}{\sqrt 2}+\frac{\eta}{\sqrt 6}&K^0\\
			K^-&\bar K^0&-\sqrt{\frac{2}{3}}\eta
		\end{pmatrix},
	\end{align}
	where we identify the $\eta$ with the octet $\eta_8$
	and assume the singlet coupling to be zero.
	 The $D$ and $F$ couplings appearing in the Lagrangian in Eq.~\eqref{appen:eq:pseudoscalar} have relation with two independent couplings $g_1$ and $g_2$ defined in Ref.~\cite{deSwart:1963pdg}, such as 
	\begin{align}
		D=\frac{\sqrt{30}}{40}g_1, \ F=\frac{\sqrt{6}}{24}g_2,
	\end{align}
	and these can be expressed with another two independent parameters, i.e. the nucleon-nucleon-pion coupling $g_{NN\pi}=D+F$ and mixing parameter $\alpha_P=F/(D+F)$ [$\alpha_P$ connects the $g_{NN\pi}$ to other couplings], where we adopt the notations used in Ref.~\cite{deSwart:1963pdg}. With Lagrangian in Eq.~\eqref{appen:eq:pseudoscalar}, we can reproduce the relation of relevant coupling constants given in Refs.~\cite{deSwart:1963pdg,Ronchen:2012eg}. Similarly, the Lagrangian for octet baryon and nonet vector meson interaction takes the form 
	\begin{align}
	\mathcal{L}_{\mathcal{B} \mathcal{B} V}&=-\sqrt 2D'\langle\bar{\mathcal{B}} \gamma_\mu [\tilde{\mathbb{V}}^\mu_1,\mathcal{B}]_+\rangle-\sqrt 2F'\langle\bar{\mathcal{B}}\gamma_\mu [\tilde{\mathbb{V}}^\mu_1,\mathcal{B}]_-\rangle\notag\\
	&+\frac{\sqrt 2D''}{2m_{\mathcal{B}}}\langle\bar{\mathcal{B}} \sigma_{\mu\nu}\partial^\nu [\tilde{\mathbb{V}}^\mu_2,\mathcal{B}]_+\rangle+\frac{\sqrt 2F''}{2m_{\mathcal{B}}}\langle\bar{\mathcal{B}}\sigma_{\mu\nu}\partial^\nu[\tilde{\mathbb{V}}^\mu_2,\mathcal{B}]_-\rangle,\label{appen:eq:vector}
	\end{align}  
	where $\tilde{\mathbb{V}}_i$ is the nonet vector meson matrix, in which octet $\omega_8$ and singlet $\omega_1$ states are not mixed, $D'(D'')$ and $F'(F'')$ are the two independent coupling for vector (tensor) currents.	Matrix form of $\tilde{\mathbb{V}}_i$ is  
\begin{align}
	\tilde{\mathbb{V}}_i=
	\begin{pmatrix}
		\frac{\rho^0}{\sqrt 2}+\frac{\omega_8}{\sqrt 6}&\rho^+&K^{*+}\\
		\rho^-&-\frac{\rho^0}{\sqrt 2}+\frac{\omega_8}{\sqrt 6}&K^{*0}\\
		K^{*-}&\bar K^{*0}&-\sqrt{\frac{2}{3}}\omega_8
	\end{pmatrix}+g'_i\begin{pmatrix}
		\omega_1&0&0\\
		0&\omega_1&0\\
		0&0&\omega_1
	\end{pmatrix},
\end{align}
where $i=1,2$, $g'_1$ and $g'_2$ describe the couplings of the singlet vector meson $\omega_1$ via vector and tensor currents. The couplings of the physical $\omega$ and $\phi$ are obtained by assuming ideal mixing of $\omega_8$ and $\omega_1$\cite{ParticleDataGroup:2020ssz}
\begin{align}
\begin{pmatrix}
	\omega_8\\\omega_1
\end{pmatrix}=
\begin{pmatrix}
	\sqrt{\frac{1}{3}}&\sqrt{\frac{2}{3}}\\\sqrt{\frac{2}{3}}&-\sqrt{\frac{1}{3}}
\end{pmatrix}
\begin{pmatrix}
	\omega\\\phi
\end{pmatrix}.
\end{align}
Furthermore, we assume that the $\phi$ meson does not couple to the nucleon (Okubo-Zweig-Iizuka rule) to fix the singlet coupling constants to be $g'_1=(3F'-D')/(2\sqrt 3 D')$ and $g'_2=(3F''-D'')/(2\sqrt 3 D'')$. The vertex of nucleon-nucleon-$\rho$ meson interaction is ~\cite{Ronchen:2012eg}
\begin{align}
	\mathcal{L}_{NN\rho}=-g_{NN\rho}\bar\psi_N\{ \gamma_\mu-\frac{\kappa_{N N \rho}}{2 m_N}\sigma_{\mu\nu}\partial^\nu\}\vec\tau\cdot \vec\rho^\mu\psi_N.
\end{align}
 After expending the Lagrangian in Eq.~\eqref{appen:eq:vector}, couplings of vector and tensor currents ($g_{NN\rho}$ and $f_{NN\rho}=g_{NN\rho}\kappa_{NN\rho}$) are expressed in terms of  $D'$, $F'$, $D''$ and $F''$, such as 
\begin{align}
	g_{NN\rho}=D'+F',\ f_{NN\rho}=D''+F'', .
\end{align}
For the nonet vector meson, we have two mixing parameters, $\alpha_V=F'/(F'+D')$ and $\alpha_V'=F''/(F''+D'')=1/4$, and the value of the later one is fixed from the hypothesis $f_{NN\omega}=0$~\cite{Ronchen:2012eg}. With knowledge above, we can also reproduce the relation of couplings relevant to octet baryon and vector meson interaction in Ref.~\cite{Ronchen:2012eg}, and expressing the couplings for other octet baryon ($\Lambda$ and $\Sigma$) in terms of nucleon's as the manner in Table~\ref{tab:coupling} is obvious.  \\

\section{Relative sign of the coupling constants}\label{appen:phase}
The sign of the coupling constants can be fixed by the quark model. The procedure is to calculate the effective vertices twice at quark level and at  hadronic level, and to equate them (only a rough estimation can be made on the strength of the couplings, it does however determine their signs)~\cite{Riska:2000gd}. The effective Lagrangian depicting the interactions of the light constituent $u,~d$ quark fields $\psi_q$ with $\pi,\rho$ and $\sigma$ mesons can be written as      
\begin{align}
	\mathcal{L}_q&=-\frac{g_{qq\pi}}{m_\pi}\bar \psi_q \gamma^5\gamma^\mu(\partial_\mu  \vec \tau\cdot \vec \pi) \psi_q\notag\\
	 &-g_{qq\rho}\bar\psi_q\{ \gamma_\mu-\frac{\kappa_{qq\rho}}{2 m_q}\sigma_{\mu\nu}\partial^\nu\}\vec\tau\cdot \vec\rho^\mu\psi_q\notag\\
	 &-g_{qq\sigma}\bar \psi_q \sigma \psi_q,\label{appen:eq-lagqq}
\end{align}
where $\vec \tau$ is the Pauli matrix, representing the isospin. For convenience, the currents in Eq.~\eqref{appen:eq-lagqq} can be written as 
\begin{align}
	j_{qq\pi}^{\mu,a}&=-\frac{g_{qq\pi}}{m_\pi}\bar \psi_q\gamma^5\gamma^\mu \tau^a\psi_q,\label{appen:eq-qqpiCurr}\\
	j_{qq\rho}^{\mu,a}&=-g_{qq\rho}\bar \psi_q\gamma^\mu \tau^a\psi_q,\\
	t_{qq\rho}^{\mu\nu,a}&=\frac{g_{qq\rho}\kappa_{qq\rho}}{2m_q}\bar \psi_q\sigma^{\mu\nu} \tau^a\psi_q,\\
	j_{qq\sigma}&=-g_{qq\sigma}\bar \psi_q\psi_q,\label{appen:eq-qqsigmaCurr}
\end{align}
where $j_{qq\pi}^{\mu,a}$, $j_{qq\rho}^{\mu,a}$, $t_{qq\rho}^{\mu\nu,a}$,  $j_{qq\sigma}$ couple with $\partial_\mu \pi^a$, $\rho^a_\mu$, $\partial_\nu \rho^a_\mu$ and $\sigma$, respectively. The spin-flavor wave functions of the proton and $\bar D^{*0}$ meson with $s_z$ (third component of the spin) can be written as
\begin{widetext}
\begin{align}
	|p,s_z\rangle&=\left [ \frac{1}{\sqrt 2}\mathcal{C}_{\frac{1}{2},s_{z1};\frac{1}{2},s_{z2}}^{1,s_{z12}}\mathcal{C}_{1,s_{z12};\frac{1}{2},s_{z3}}^{\frac{1}{2},s_z}   \mathcal{C}_{\frac{1}{2},I_{z1};\frac{1}{2},I_{z2}}^{1,I_{z12}}\mathcal{C}_{1,I_{z12};\frac{1}{2},I_{z3}}^{\frac{1}{2},\frac{1}{2}}+\frac{1}{\sqrt 2}\mathcal{C}_{\frac{1}{2},s_{z1};\frac{1}{2},s_{z2}}^{0,0}\mathcal{C}_{\frac{1}{2},I_{z1};\frac{1}{2},I_{z2}}^{0,0}\delta_{s_{z},s_{z3}}\delta_{\frac{1}{2},I_{z3}}\right ] |s_{z1},s_{z2},s_{z3} \rangle |I_{z1},I_{z2},I_{z3} \rangle,\\
   |\bar D^{*0},s_z\rangle&=\mathcal{C}_{\frac{1}{2},s_{z1};\frac{1}{2},s_{z2}}^{1,s_{z}}\delta_{\frac{1}{2},I_{z1}}\delta_{I_{z2},0}|s_{z1},s_{z2} \rangle |I_{z1},I_{z2} \rangle,
\end{align}      
\end{widetext}
where $s_{zi}$($I_{zi}$) is the third component of the spin(isospin) for the $i$th quark, $\mathcal{C}_{A,a;B,b}^{C,c}$ represents the Clebsch-Gordan coefficient, $\delta_{a,b}$ is a Kronecker delta function, the quantum numbers [$s_{z1}$, $s_{z2}$, $s_{z12}$, $s_{z3}$, $I_{z1}$, $I_{z2}$, $I_{z12}$, $I_{z3}$] should be summed. With the proton wave function above, we can calculate the matrix elements of the currents in Eqs.~\eqref{appen:eq-qqpiCurr}-\eqref{appen:eq-qqsigmaCurr} for a proton with spin up,  
\begin{subequations}\label{appen:eq-quark-proton}
	\begin{align}
	\langle p,\frac{1}{2}|\sum\limits_{q=1}^3j_{qq\pi}^{3,3}| p,\frac{1}{2} \rangle&=\frac{5}{3}\frac{g_{qq\pi}}{m_\pi},\label{appen:eq-quark-NNpi}\\
	\langle p,\frac{1}{2}|\sum\limits_{q=1}^3j_{qq\rho}^{0,3}| p,\frac{1}{2} \rangle&=-g_{qq\rho},\\
	\langle p,\frac{1}{2}|\sum\limits_{q=1}^3t_{qq\rho}^{21,3}| p,\frac{1}{2} \rangle&=-\frac{5}{6}\frac{g_{qq\rho}\kappa_{qq\rho}}{m_q},\\
	\langle p,\frac{1}{2}|\sum\limits_{q=1}^3j_{qq\sigma}| p,\frac{1}{2} \rangle&=-3 g_{qq\sigma},
\end{align}    
\end{subequations}
where $\sum\limits_{q=1}^3$ represents the sum over the three quarks in the proton. Similarly, for the $\bar D^{*0}$ meson with $s_z=1$, the matrix elements of the currents in Eqs.~\eqref{appen:eq-qqpiCurr}-\eqref{appen:eq-qqsigmaCurr} are
\begin{subequations}\label{appen:eq-quark-D}
\begin{align}
	\langle \bar D^{*0},1|j_{qq\pi}^{3,3}| \bar D^{*0},1 \rangle&=\frac{ g_{qq\pi}}{m_\pi},\label{appen:eq-quark-DDpi}\\
	\langle \bar D^{*0},1|j_{qq\rho}^{0,3}| \bar D^{*0},1 \rangle&=- g_{qq\rho},\\
	\langle \bar D^{*0},1|t_{qq\rho}^{21,3}| \bar D^{*0},1 \rangle&=-\frac{g_{qq\rho}\kappa_{qq\rho}}{2m_q},\\
	\langle \bar D^{*0},1|j_{qq\sigma}| \bar D^{*0},1 \rangle&=- g_{qq\sigma}.
\end{align}  
\end{subequations}

Now, we calculate the same currents at hadronic level. Considering the isospin parts in the Lagrangian in Eqs~\eqref{lag:BBsigma}-\eqref{lag:BBP}, the effective vertices for the interaction of nucleon with $\pi,\rho$ and $\sigma$ mesons can be written as 
\begin{align}
	\mathcal{L}_{NN\pi}&=-\frac{g_{NN\pi}}{m_\pi}\bar\psi_N \gamma^5\gamma^\mu (\partial_\mu  \vec \tau\cdot \vec \pi)\psi_N,\\
	\mathcal{L}_{NN\rho}&=-g_{NN\rho}\bar\psi_N\{ \gamma_\mu-\frac{\kappa_{N N \rho}}{2 m_N}\sigma_{\mu\nu}\partial^\nu\}\vec\tau\cdot \vec\rho^\mu\psi_N,\\
	\mathcal{L}_{NN\sigma}&=-g_{NN\sigma}\bar\psi_N\sigma\psi_N.
\end{align}
We can also define similar currents for nucleon vertices,
\begin{align}
	j_{NN\pi}^{\mu,a}&=-\frac{g_{NN\pi}}{m_\pi}\bar \psi_N\gamma^5\gamma^\mu \tau^a\psi_N,\label{appen:eq-NNpiCurr}\\
	j_{NN\rho}^{\mu,a}&=-g_{NN\rho}\bar \psi_N\gamma^\mu \tau^a\psi_N,\\
	t_{NN\rho}^{\mu\nu,a}&=\frac{g_{NN\rho}\kappa_{NN\rho}}{2m_N}\bar \psi_N\sigma^{\mu\nu} \tau^a\psi_N,\\
	j_{NN\sigma}&=-g_{NN\sigma}\bar \psi_N\psi_N.\label{appen:eq-NNsigmaCurr}
\end{align}
For the proton with spin up, the matrix elements of them calculated at hadronic level are
\begin{subequations}\label{appen:eq-hadron-proton}
	\begin{align}
		\langle p,\frac{1}{2}|j_{NN\pi}^{3,3}| p,\frac{1}{2} \rangle&=\frac{ g_{NN\pi}}{m_\pi},\label{appen:eq-hadron-NNpi}\\
		\langle p,\frac{1}{2}|j_{NN\rho}^{0,3}| p,\frac{1}{2} \rangle&= g_{NN\rho},\\
		\langle p,\frac{1}{2}|t_{NN\rho}^{21,3}| p,\frac{1}{2} \rangle&=- \frac{g_{NN\rho}\kappa_{NN\rho}}{2m_N},\\
		\langle p,\frac{1}{2}|j_{NN\sigma}| p,\frac{1}{2} \rangle&=-g_{NN\sigma}.
	\end{align}
\end{subequations}
After expanding the Lagrangian in Eqs.~\eqref{lag:ppsigma}-\eqref{lag:ppP} in flavor space, interaction vertices of the $\bar D^*$ meson with the $\pi,\rho$ and $\sigma$ mesons are
\begin{align}
	\mathcal{L}_{\bar D^*\bar D^*\pi}&=i\frac{2g}{\sqrt 2f_\pi}\varepsilon^{\alpha\mu\nu\kappa}v_\kappa \bar D^{*\dagger}_\mu (\partial_\alpha  \vec \tau\cdot \vec \pi)\bar D^*_\nu,\\
	\mathcal{L}_{\bar D^*\bar D^*\rho}&=-\beta g_V \bar D^{*\dagger}_\mu v^\alpha \vec\tau\cdot \vec\rho_\alpha \bar D^{*\mu}\notag\\
	&-i2\lambda g_V  \bar D^{*\mu\dagger}(\partial_\mu \vec\tau\cdot\vec\rho_\nu-\partial_\nu \vec\tau\cdot\vec\rho_\mu)\bar D^{*\nu},\\
	\mathcal{L}_{\bar D^*\bar D^*\sigma}&=2g_S\bar D^{*\dagger}_\mu \sigma \bar D^{*\mu},
\end{align}
where $\bar D^*$ is a scaled filed satisfying $\langle 0|\bar{D}^*_\mu|\bar c q(1^-)\rangle=\epsilon_\mu\sqrt{M_{\bar{D}^*}}$ as $\mathcal{P}^*$ field, and it is written in the isospin space as $\bar D^*=(\bar D^{*0},D^{*-})^T$. The similar currents for $\pi,\rho$ and $\sigma$ mesons can be written as 
\begin{align}
	j_{\bar D^*\bar D^*\pi}^{\alpha,a}&=i\frac{2g}{\sqrt 2f_\pi}\varepsilon^{\alpha\mu\nu\kappa}v_\kappa \bar D^{*\dagger}_\mu\tau^a\bar D^*_\nu,\label{appen:eq-DDpiCurr}\\
	j_{\bar D^*\bar D^*\rho}^{\alpha,a}&=-\beta g_V \bar D^{*\dagger}_\mu v^\alpha\bar D^{*\mu},\\
	t_{\bar D^*\bar D^*\rho}^{\mu\nu,a}&=i2\lambda g_V  (\bar D^{*\mu\dagger}\tau^a\bar D^{*\nu}-\bar D^{*\nu\dagger}\tau^a\bar D^{*\mu}),\\
	j_{\bar D^*\bar D^*\sigma}&=2g_S\bar D^{*\dagger}_\mu\bar D^{*\mu}.\label{appen:eq-DDsigmaCurr}
\end{align}
For $\bar D^{*0}$ meson with $s_z=1$, the matrix elements of these currents at hadronic level are 
\begin{subequations}\label{appen:eq-hadron-D}
\begin{align}
	\langle \bar D^{*0},1|j_{\bar D^*\bar D^*\pi}^{3,3}| \bar D^{*0},1 \rangle&=-\frac{\sqrt 2m_{\bar D^*} g}{f_\pi},\label{appen:eq-hadron-DDpi}\\
	\langle \bar D^{*0},1|j_{\bar D^*\bar D^*\rho}^{0,3}| \bar D^{*0},1 \rangle&= m_{\bar D^*}\beta g_V,\\
	\langle \bar D^{*0},1|t_{\bar D^*\bar D^*\rho}^{21,3}| \bar D^{*0},1 \rangle&=-2m_{\bar D^*}\lambda g_V,\\
	\langle \bar D^{*0},1|j_{\bar D^*\bar D^*\sigma}| \bar D^{*0},1 \rangle&=-2m_{\bar D^*}g_S.
\end{align} 
\end{subequations}

With the assumption that the currents calculated at the quark level [Eqs.~\eqref{appen:eq-quark-proton},~\eqref{appen:eq-quark-D}] and hadronic level [Eqs.~\eqref{appen:eq-hadron-proton},~\eqref{appen:eq-hadron-D}] are consistent in sign, we can determine the relative sign between $g_{NN\pi}$ and $g$, $g_{NN\rho}$ and $\beta$, $\kappa_{NN\rho}$ and $\lambda$, $g_{NN\sigma}$ and $g_S$. For instance, the relative sign between the currents in Eq.~\eqref{appen:eq-quark-NNpi} and Eq.~\eqref{appen:eq-hadron-NNpi} which couple to the $\pi^3$ with $k_3$ (third component of the $\pi^3$ momentum) requires that $g_{qq\pi}$ and $g_{NN\pi}$ have the same sign, while the relative sign between the currents in Eq.~\eqref{appen:eq-quark-DDpi} and Eq.~\eqref{appen:eq-hadron-DDpi} indicates that the signs of $g_{qq\pi}$ and $g$ are opposite. Thus, we can determine the relative sign between $g_{NN\pi}$ and $g$, and they are opposite to each other. In this way, we can determine the relative sign between $g_{NN\rho}$ and $\beta$, $\kappa_{NN\rho}$ and $\lambda$, $g_{NN\sigma}$ and $g_S$, that is, both of them have the same sign (positive or negative). Also, the signs of other octet baryon couplings $g_{\mathcal{B}\mathcal{B}V}$ and $g_{\mathcal{B}\mathcal{B}P}$ are fixed with respect to that of nucleons in the SU(3) flavor symmetry. For the scalar couplings, we assume that the signs of $g_{\Sigma\Sigma\sigma}$ and $g_{\Lambda\Lambda\sigma}$ are the same as $g_{NN\sigma}$.

   \end{appendices}
\bibliographystyle{apsrev4-1}
\bibliography{reference.bib}
\end{document}